\documentstyle[12pt]{article}

\newcommand{\noi}{\noindent}
\newcommand{\hsp}{\hspace{5.7mm}}
\newcommand{\mini}{\hspace{.3mm}}      
\newcommand{\mmini}{\hspace{-.4mm}}    
\newcommand{\mimini}{\hspace{-.3mm}}
\newcommand{\mA}{\hspace{-1mm}}

\let\ssection=\section
\renewcommand{\section}{\setcounter{equation}{0}\ssection}

\newfont{\BBB}{msbm10 scaled\magstephalf}
\newcommand{\BB}[1]{\mbox{\BBB #1}}
\newcommand{\BBn}[2]{\mbox{{\BBB #1}$^{#2}$}}
\newcommand{\BBd}[2]{\mbox{{\BBB #1}$_{#2}$}}

\newcommand{\la}{<\!}
\newcommand{\ra}{\!>\:}
\newcommand{\las}{<}
\newcommand{\ras}{>}

\newcommand{\be}{\begin{equation}}
\newcommand{\ee}{\end{equation} \vspace{0mm}}
\newcommand{\ben}{$$}       
\newcommand{\een}{$$ \vspace{0mm}}
\newcommand{\bea}{\begin{eqnarray}}
\newcommand{\eea}{\end{eqnarray} \vspace{0mm}}
\newcommand{\ba}{\begin{array}{c}}
\newcommand{\ea}{\end{array}}

\newcommand{\cfr}[1]{(\ref{#1})}
\newcommand{\cfrs}[2]{(\ref{#1}) and~(\ref{#2})}

\newcommand{\hah}{\hsp\mbox{and}\hsp}

\newcommand{\hfa}{\hsp\mbox{for all }\,}
\newcommand{\hfs}{\hsp\mbox{for some }\,}
\newcommand{\hwh}{\hsp\mbox{with}\hsp}
\newcommand{\hfh}{\hsp\mbox{for}\hsp}

\newcommand{\vav}{\vspace{-4.5mm}\noi and \vspace{-3mm}}
\newcommand{\vwv}{\vspace{-4.5mm}\noi where \vspace{-3mm}}
\newcommand{\Lmini}{$L\mini$-minimal}
\newcommand{\Lminiy}{$L\mini$-minimality}

\newcommand{\sca}[2]{\la #1 , #2 \ra}
\newcommand{\avecd}[1]{#1_{1} , \ldots , #1_{N} }
\newcommand{\avecu}[1]{#1^{1} , \ldots , #1^{N} }
\newcommand{\diix}{d^{\mini 2}\mmini x}
\newcommand{\BQ}{{\cal B}_{\!\Q}}
\newcommand{\BQs}{\raisebox{-.5mm}{\mbox{${\textstyle\cal
   B}_{\scriptscriptstyle\bf Q}$}}}

\newcommand{\Sp}{{\Sigma}_{p}}
\newcommand{\Sm}{{\cal S}_{p}}
\newcommand{\Z}{\BB{Z}}
\newcommand{\Q}{{\bf Q}}
\newcommand{\q}{{\bf q}}
\newcommand{\e}{{\bf e}}
\newcommand{\f}{{\bf f}}
\newcommand{\n}{{\bf n}}

\newcommand{\G}{\Gamma}
\newcommand{\Gs}{\Gamma^\ast}
\newcommand{\Gp}{\Gamma_{\!phys}}
\newcommand{\Hp}{{\cal H}_p}

\newcommand{\Sigp}{\Sigma_p}

\newcommand{\D}{\Delta}

\newcommand{\LMg}{L_{max}(\G,\Q)}
\newcommand{\LM}{L_{max}}
\newcommand{\Lmg}{L_{min}(\G,\Q)}
\newcommand{\Lm}{L_{min}}
\newcommand{\Ls}{\ell_{*}}
\newcommand{\Ns}{N_{*}}
\newcommand{\lMg}{\ell_{max}(\G,\Q)}
\newcommand{\lM}{\ell_{max}}
\newcommand{\lmg}{\ell_{min}(\G,\Q)}
\newcommand{\lm}{\ell_{min}}

\newcommand{\sH}{\sigma_{\!\scriptscriptstyle H}}
\newcommand{\nH}{n_{\mimini\scriptscriptstyle H}}
\newcommand{\dH}{d_{\mimini\scriptscriptstyle H}}
\newcommand{\RH}{R_{\scriptscriptstyle H}}
\newcommand{\RL}{R_{\scriptscriptstyle L}}

\newcommand{\es}{\mbox{\boldmath$\varepsilon$}}

\newcommand{\suhN}{$\,\widehat{su}(N)$}

\newcommand{\ui}{$u(1)$}

\newcommand{\PRL}{Phys.\ Rev.\ Lett.\ }
\newcommand{\PRB}{Phys.\ Rev.\ B }

\newcommand{\PR}{Phys.\ Rev.\ }

\newcommand{\NPB}{Nucl.\ Phys.\ B }
\newcommand{\AP}{Ann.\ Phys.\ (N.Y.) }
\newcommand{\CMP}{Commun.\ Math.\ Phys.\ }
\newcommand{\SurS}{Surf.\ Sci.\ }

\newcommand{\IJMP}{Int.\ J.\ Mod.\ Phys.\ }

\begin{document}


\thispagestyle{empty}

\begin{flushright}
{\footnotesize\tt preprint ETH-TH/95-5 \\ May 1995 \\
arch-ive/9505156}
\end{flushright}

\vspace{20mm}

\begin{center}
{\large\bf STRUCTURING THE SET OF\rule[-4mm]{0mm}{6mm} \\
INCOMPRESSIBLE \mbox{QUANTUM} HALL FLUIDS}

\vspace{15mm}

J\"urg Fr\"ohlich$^1$, Thomas Kerler$^2$,
Urban M.~Studer,$^3$\footnote{Present address: Institut f\"ur
Theoretische Physik, ETH-H\"onggerberg, 8093 Z\"urich, Switzerland}
and Emmanuel Thiran$^1$

\vspace{10mm}

$^1\,${\small Institut f\"ur Theoretische Physik, ETH-H\"onggerberg,
8093 Z\"urich, Switzerland} \\
$^2\,${\small Department of Mathematics, Harvard University,
Cambridge, MA 02138, USA} \\
$^3\,${\small Instituut voor Theoretische Fysica, Katholieke
Universiteit Leuven, 3001 Leuven, Belgium}
\end{center}

\vspace{23mm}

\begin{list}
{}{\setlength{\leftmargin}{0mm}\setlength{\rightmargin}{0mm}
\setlength{\parsep}{0mm}\setlength{\listparindent}{8.6mm}}
\item
\small
{\bf Abstract:} A classification of incompressible quantum Hall
fluids in terms of integral lattices and arithmetical invariants
thereof is proposed. This classification enables us to
characterize the plateau values of the Hall conductivity $\sH$
in the interval $\,(0,1]\,$ (in units where $\,e^2/h=1$)
corresponding to ``stable'' incompressible quantum Hall fluids. A
bijection, called shift map, between classes of stable
incompressible quantum Hall fluids corresponding to plateaux of
$\sH$ in the intervals $\,[1/(2\mini p+1),1/(2\mini p-1)\mini)\,$
and $\,[1/(2\mini q+1),1/(2\mini q-1)\mini)$, respectively, is
constructed, with $\,p,q=1,2,3, (\ldots),\ p\neq q$.

Our theoretical results are carefully compared to experimental
data, and various predictions and experimental implications of our
theory are discussed.
\end{list}


\newpage
\setcounter{page}{4}
\begin{flushleft}
\vspace*{3mm}
\begin{picture}(10,8)(-.4,0)

\put(4.74,0){\makebox(0,0)[l]
  {$\hspace*{.4mm}\cdot\hspace{1.2mm}{\bf 9\over 19}$}}

\put(4.71,.8){\makebox(0,0)[l]
  {$\circ\hspace{.6mm}{\bf 8\over 17}$}}
\put(5.29,.8){\makebox(0,0)[l]
  {$\circ\hspace{.6mm}{\bf 9\over 17}$}}
\put(5.88,.8){\makebox(0,0)[l]
  {$\hspace*{.4mm}\cdot\hspace{1.2mm}{\bf 10\over 17}$}}

\put(2.67,1.6){\makebox(0,0)[l]
  {$\circ\hspace{.7mm}{\bf 4\over 15}$}}
\put(4.67,1.6){\makebox(0,0)[l]
  {$\circ\hspace{.7mm}{\bf 7\over 15}$}}
\put(5.33,1.6){\makebox(0,0)[l]
  {$\circ\hspace{.7mm}{\bf 8\over 15}$}}

\put(2.3,2.4){\makebox(0,0)[l]
  {$\circ\hspace{.7mm}{\bf 3\over 13}$}}
\put(3.07,2.4){\makebox(0,0)[l]
  {$\hspace*{.4mm}\cdot\hspace{1.2mm}{\bf 4\over 13}$}}
\put(4.62,2.4){\makebox(0,0)[l]
  {$\bullet\hspace{.7mm}{\bf 6\over 13}$}}
\put(5.38,2.4){\makebox(0,0)[l]
  {$\bullet\hspace{.7mm}{\bf 7\over 13}$}}
\put(6.15,2.4){\makebox(0,0)[l]
  {$\bullet\hspace{.7mm}{\bf 8\over 13}$}}
\put(6.92,2.4){\makebox(0,0)[l]
  {$\circ\hspace{.7mm}{\bf 9\over 13}$}}

\put(1.82,3.2){\makebox(0,0)[l]
  {$\circ\hspace{.7mm}{\bf 2\over 11}$}}
\put(2.73,3.2){\makebox(0,0)[l]
  {$\bullet\hspace{.7mm}{\bf 3\over 11}$}}
\put(3.64,3.2){\makebox(0,0)[l]
  {$\hspace*{.4mm}\cdot\hspace{1.2mm}{\bf 4\over 11}$}}
\put(4.54,3.2){\makebox(0,0)[l]
  {$\bullet\hspace{.5mm}{\bf 5\over 11}$}}
\put(5.45,3.2){\makebox(0,0)[l]
  {$\bullet\hspace{.7mm}{\bf 6\over 11}$}}
\put(6.36,3.2){\makebox(0,0)[l]
  {$\circ\hspace{.7mm}{\bf 7\over 11}$}}
\put(7.27,3.2){\makebox(0,0)[l]
  {$\bullet\hspace{.7mm}{\bf 8\over 11}$}}

\put(2.22,4){\makebox(0,0)[l]
  {$\bullet\hspace{.7mm}{\bf 2\over 9}$}}
\put(4.44,4){\makebox(0,0)[l]
  {$\bullet\hspace{.7mm}{\bf 4\over 9}$}}
\put(5.56,4){\makebox(0,0)[l]
  {$\bullet\hspace{.7mm}{\bf 5\over 9}$}}

\put(1.43,4.8){\makebox(0,0)[l]
  {$\circ\hspace{.7mm}{\bf 1\over 7}$}}
\put(2.86,4.8){\makebox(0,0)[l]
  {$\bullet\hspace{.7mm}{\bf 2\over 7}$}}
\put(4.29,4.8){\makebox(0,0)[l]
  {$\bullet\hspace{.7mm}{\bf 3\over 7}$}}
\put(5.71,4.8){\makebox(0,0)[l]
  {$\bullet\hspace{.7mm}{\bf 4\over 7}$}}
\put(7.14,4.8){\makebox(0,0)[l]
  {$\bullet\hspace{.7mm}{\bf 5\over 7}\:${\scriptsize\em
  (B-p\hspace{-.2mm}) }}}

\put(2,5.6){\makebox(0,0)[l]
  {$\bullet\hspace{.7mm}{\bf 1\over 5}$}}
\put(4,5.6){\makebox(0,0)[l]
  {$\bullet\hspace{.7mm}{\bf 2\over 5}\:${\scriptsize\em
  (B-p\hspace{-.2mm}) }}}
\put(6,5.6){\makebox(0,0)[l]
  {$\bullet\hspace{.7mm}{\bf 3\over 5}\:${\scriptsize\em B-p}}}
\put(8,5.6){\makebox(0,0)[l]
  {$\bullet\hspace{.7mm}{\bf 4\over 5}$}}

\put(3.33,6.4){\makebox(0,0)[l]
  {$\bullet\hspace{.7mm}{\bf 1\over 3}$}}
\put(6.67,6.4){\makebox(0,0)[l]
  {$\bullet\hspace{.7mm}{\bf 2\over 3}\:${\scriptsize\em B/n-p}}}

\put(10,7.2){\makebox(0,0)[l]
  {$\bullet\hspace{1mm}{\bf 1}$}}

\put(.73,7.2){\makebox(0,0)[r]{\small$\dH=1$}}
\put(.73,6.4){\makebox(0,0)[r]{\small$3$}}
\put(.73,5.6){\makebox(0,0)[r]{\small$5$}}
\put(.73,4.8){\makebox(0,0)[r]{\small$7$}}
\put(.73,4){\makebox(0,0)[r]{\small$9$}}
\put(.73,3.2){\makebox(0,0)[r]{\small$11$}}
\put(.73,2.4){\makebox(0,0)[r]{\small$13$}}
\put(.73,1.6){\makebox(0,0)[r]{\small$15$}}
\put(.73,.8){\makebox(0,0)[r]{\small$17$}}
\put(.73,0){\makebox(0,0)[r]{\small$19$}}

\thinlines
\put(-.5,-.45){\hspace{1mm}\line(1,0){11}}
\put(-.5,7.6){\hspace{1mm}\line(1,0){11}}

\multiput(0,7.6)(0,-0.1){3}{\hspace{1mm}\line(0,-1){0.04}} 
\multiput(0,7)(0,-0.1){74}{\hspace{1mm}\line(0,-1){0.04}}
\put(0,-.4){\hspace{1mm}\line(0,-1){0.1}}

\multiput(1.43,7.6)(0,-0.1){28}{\hspace{1mm}\line(0,-1){0.04}} 
\multiput(1.43,4.7)(0,-0.1){51}{\hspace{1mm}\line(0,-1){0.04}}
\put(1.43,-.4){\hspace{1mm}\line(0,-1){0.1}}

\multiput(1.67,7.6)(0,-0.2){13}{\hspace{1mm}\line(0,-1){0.1}} 
\put(1.67,4.56){\hspace{1mm}\line(0,-1){0.06}}
\multiput(1.67,4.4)(0,-0.2){25}{\hspace{1mm}\line(0,-1){0.1}}

\multiput(2,7.6)(0,-0.1){20}{\hspace{1mm}\line(0,-1){0.04}} 
\multiput(2,5.5)(0,-0.1){21}{\hspace{1mm}\line(0,-1){0.04}}
\multiput(2,2.9)(0,-0.1){33}{\hspace{1mm}\line(0,-1){0.04}}
\put(2,-.4){\hspace{1mm}\line(0,-1){0.1}}

\multiput(2.5,7.6)(0,-0.2){17}{\hspace{1mm}\line(0,-1){0.1}} 
\put(2.5,3.76){\hspace{1mm}\line(0,-1){0.06}}
\multiput(2.5,3.6)(0,-0.2){5}{\hspace{1mm}\line(0,-1){0.1}}
\put(2.5,2.16){\hspace{1mm}\line(0,-1){0.06}}
\multiput(2.5,2)(0,-0.2){13}{\hspace{1mm}\line(0,-1){0.1}}

\multiput(3.33,7.6)(0,-0.1){12}{\hspace{1mm}\line(0,-1){0.04}} 
\multiput(3.33,6.3)(0,-0.1){37}{\hspace{1mm}\line(0,-1){0.04}}
\multiput(3.33,2.1)(0,-0.1){25}{\hspace{1mm}\line(0,-1){0.04}}
\put(3.33,-.4){\hspace{1mm}\line(0,-1){0.1}}

\multiput(5,7.6)(0,-0.2){25}{\hspace{1mm}\line(0,-1){0.1}} 
\put(5,2.16){\hspace{1mm}\line(0,-1){0.06}}
\put(5,2){\hspace{1mm}\line(0,-1){0.1}}
\put(5,1.36){\hspace{1mm}\line(0,-1){0.06}}
\put(5,1.2){\hspace{1mm}\line(0,-1){0.1}}
\put(5,.56){\hspace{1mm}\line(0,-1){0.06}}
\put(5,.4){\hspace{1mm}\line(0,-1){0.1}}
\put(5,-.24){\hspace{1mm}\line(0,-1){0.07}}
\put(5,-.4){\hspace{1mm}\line(0,-1){0.1}}

\multiput(10,7.6)(0,-0.1){4}{\hspace{1mm}\line(0,-1){0.04}} 
\multiput(10,7.1)(0,-0.1){75}{\hspace{1mm}\line(0,-1){0.04}}
\put(10,-.4){\hspace{1mm}\line(0,-1){0.1}}

\thicklines
\put(0,-.6){\hspace{1mm}\makebox(0,0)[t]{$0$}}
\put(.71,-.66){\hspace{1mm}\makebox(0,0)[t]{$\cdots$}}
\put(-.26,-1){\mbox{$\underbrace{\hspace*{19mm}}_{
  \renewcommand{\arraystretch}{.7} \ba \mbox{\small\em Wigner
  crystal} \\ \mbox{\small\em or carrier} \\ \mbox{\small\em
  freeze-out}\ea}$}}
\put(1.43,-.54){\hspace{1mm}\makebox(0,0)[t]{$1\over 7$}}
\put(1.43,-.41){\hspace{1mm}\line(1,0){.24}}
\put(1.67,-.54){\hspace{1mm}\makebox(0,0)[t]{$1\over 6$}}

\put(2,-.54){\hspace{1mm}\makebox(0,0)[t]{$1\over 5$}}
\put(2,-.41){\hspace{1mm}\line(1,0){.5}}
\put(2.5,-.54){\hspace{1mm}\makebox(0,0)[t]{$1\over 4$}}
\put(1.76,-1.5){\mbox{$\renewcommand{\arraystretch}{.7} \ba \uparrow
  \rule[-2mm]{0mm}{5mm} \\ \mbox{\small\em Fermi liquid} \\
  \mbox{\small\em behaviour}\ea$}}

\put(3.33,-.54){\hspace{1mm}\makebox(0,0)[t]{$1\over 3$}}
\put(3.33,-.41){\hspace{1mm}\line(1,0){1.67}}
\put(5,-.54){\hspace{1mm}\makebox(0,0)[t]{$1\over 2$}}
\put(4.26,-1.5){\mbox{$\renewcommand{\arraystretch}{.7} \ba \uparrow
  \rule[-2mm]{0mm}{5mm} \\ \mbox{\small\em Fermi liquid} \\
  \mbox{\small\em behaviour}\ea$}}

\put(10,-.6){\hspace{1mm}\makebox(0,0)[t]{$1$}}
\put(10,-.41){\hspace{1mm}\line(1,0){.5}}
\put(10.4,-.66){\hspace{1mm}\makebox(0,0)[t]{$\sH$}}
\put(8.41,-1){\mbox{$\underbrace{\hspace*{30mm}}_{
  \renewcommand{\arraystretch}{.7} \ba \mbox{\small\em domain of}Ê\\
  \mbox{\small\em attraction} \\ \mbox{\small\em of $\,\sH=1$}\ea}$}}

\put(.71,7.85){\hspace{1mm}\makebox(0,0){$\cdots$}}
\put(1.43,7.56){\hspace{1mm}\line(1,0){.24}}
\put(1.55,7.75){\hspace{.85mm}\makebox(0,0)[b]{$\Sigma_3^+$}}
\put(1.83,7.75){\hspace{2mm}\makebox(0,0)[b]{$\Sigma_3^-$}}

\put(2,7.56){\hspace{1mm}\line(1,0){.5}}
\put(2.25,7.75){\hspace{1.9mm}\makebox(0,0)[b]{$\Sigma_2^+$}}
\put(2.92,7.75){\hspace{1.4mm}\makebox(0,0)[b]{$\Sigma_2^-$}}

\put(3.33,7.56){\hspace{1mm}\line(1,0){1.67}}
\put(4.17,7.75){\hspace{1.2mm}\makebox(0,0)[b]{$\Sigma_1^+$}}
\put(7.5,7.75){\hspace{1.2mm}\makebox(0,0)[b]{$\Sigma_1^-$}}

\put(10,7.56){\hspace{1mm}\line(1,0){.5}}

\end{picture}
\end{flushleft}

\vspace{31mm}
\noi
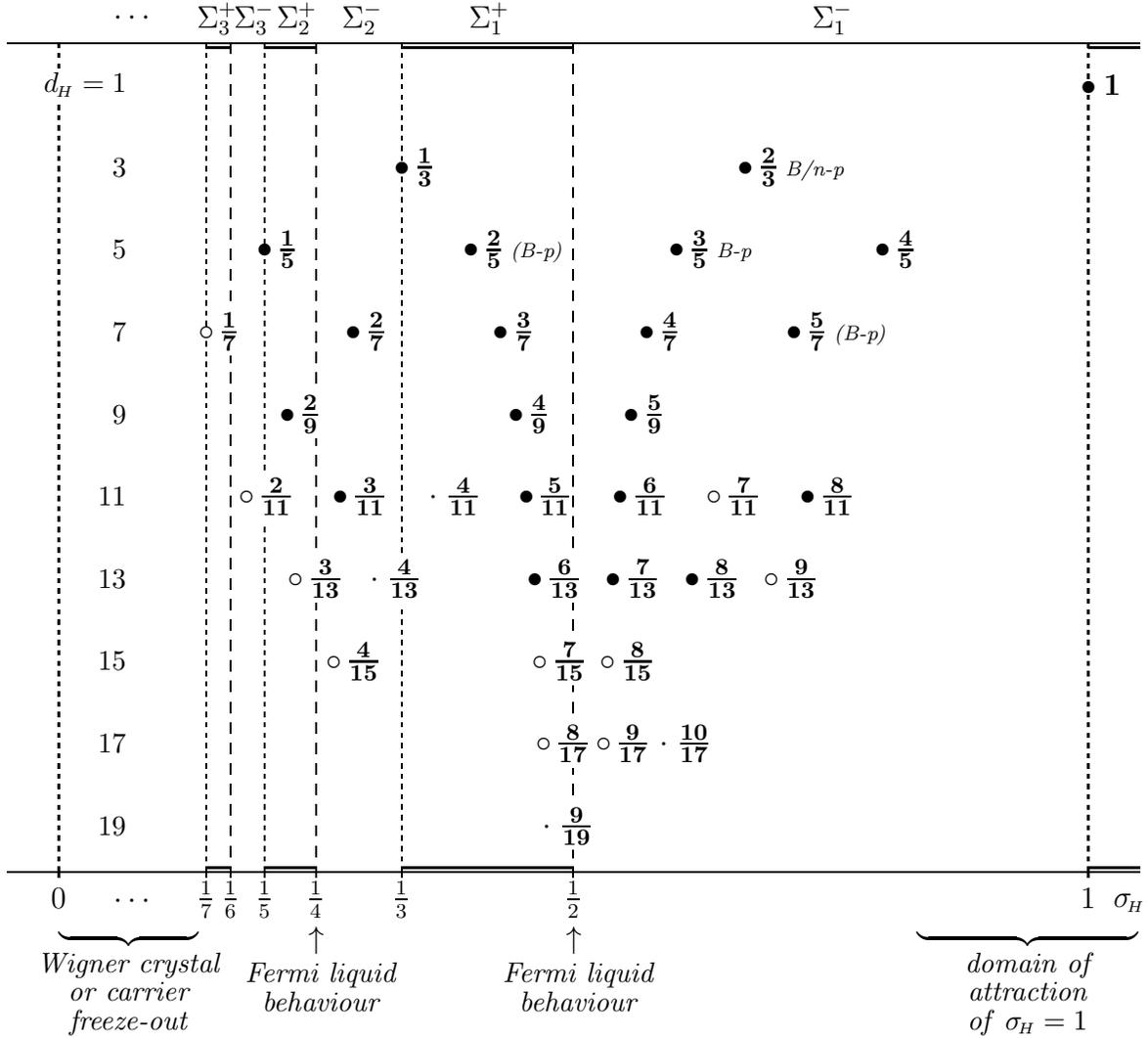
{\bf Figure 1.1. }{\em Observed Hall fractions $\,\sH=\nH/\dH\,$ in
the interval $\,0<\sH\leq 1$, and their experimental status in
single-layer quantum Hall systems.}
\rule[-4mm]{0mm}{5mm}

\noi
{\small Well established Hall fractions are indicated by
``$\,\bullet\,$''. These are fractions for which an
$R_{xx}\mini$-minimum and a plateau in $\RH$ have been clearly
observed, and the quantization accuracy of $\sH=1/\RH$ is typically
better than $0.5$\%. Fractions for which a minimum in $R_{xx}$ and
typically an inflection in $\RH$ (i.e., a minimum in ${d\mini \RH /
d\mini B_{\!c}^\perp}$, but no well developed plateau in $\RH$) have
been observed are indicated by ``$\,\circ\,$''. If there are only
very weak experimental indications or controversial data for a given
Hall fraction, the symbol ``$\,\cdot\,$'' is used. Finally, ``{\em
B/n-p\,}'' is appended to fractions at which a magnetic field
{\em\mimini(B)} and/or density {\em\mimini(n)} driven phase
transition has been observed.}


\newpage
\setcounter{page}{1}
{\small
\tableofcontents}

\vspace{15mm}


\begin{flushleft}
\section{Introduction and Summary of Contents}
\label{sI}
\end{flushleft}

In this paper, we continue our theoretical analysis of the fractional
quantum Hall (QH) effect~\cite{QHE}. We describe and explore a
{\em classification of incompressible quantum Hall fluids} (or, for
short, {\em QH fluids\mini}) in terms of pairs of an integral lattice
and a primitive vector in its dual and of arithmetical invariants of
these data. Our classification is derived form a combination of basic
physical principles and some phenomenologically well confirmed
assumptions concerning QH fluids.

Our theoretical results neatly reproduce and structure experimental
data (see Refs.~\cite{Tsui} through~\cite{Kang}) for the fractional
QH effect. They provide insight into the origin of the phase
transitions (disappearance and reappearance of plateaux when, e.g.,
the in-plane component of the external magnetic field is varied, at
a fixed value of the filling factor) that have been observed in
single-layer~\cite{Saj,Cla,Eis,Eng} and double-layer or
wide-single-quantum-well~\cite{Su} samples. We also describe the
possible structures of QH fluids corresponding to Hall fractions
corresponding to even denominators~\cite{evenSingle,evenDouble}.

The analysis presented in this paper continues the thread of
arguments initiated in Refs.~\cite{FK} through~\cite{FT}. It is a
companion of the results we have discussed in~\cite{FST}.
In~\cite{FST}, we have focussed our attention on an {\em explicit\,}
and {\em systematic\mini} classification of some {\em specific\mini}
classes of incompressible QH fluids. In the present paper, we
derive {\em general\,} mathematical {\em organizing principles\mini}
that enable us to reproduce and interpret the set of experimental
data on incompressible QH fluids and propose some new experiments.

The theoretical framework underlying our work has originally been
inspired by Halperin's analysis of edge currents~\cite{H} and is
related to ideas of Read~\cite{Read}, Stone~\cite{S} and Wen and
collaborators~\cite{Wen,WB,WZ}. The mathematical data, a pair of an
integral lattice, $\G$, and a primitive vector, $\Q$, in its
dual, $\Gs$,\,--\,called a ``{\em QH lattice\,}''\,--\,,
characterizing a universality class of (incompressible) QH fluids
become apparent when one studies the large-scale, low-frequency
properties, i.e., the physics in the scaling limit, of QH fluids. It
is convenient to describe the physics of QH fluids in the scaling
limit in terms of an effective field theory of their conserved
current densities. For {\em incompressible\mini} (dissipation-free)
QH fluids, i.e., ones with vanishing longitudinal resistance, the
effective field theory of conserved current densities is {\em
topological}, and, because of the presence of an external magnetic
field, it breaks parity and time-reversal
invariance~\cite{FZ,FSC,FSR}. It turns out that this effective field
theory is an abelian Chern-Simons theory of the vector potentials of
conserved current densities. Physical state vectors of this theory
are labelled by the points of a lattice $\Gp$ containing the lattice
$\G$ and contained in (or equal to) the dual lattice $\Gs$, and the
vector $\Q$ in $\Gs$, called ''{\em charge vector\mini}'', determines
the electric charge of a state labelled by a point in $\Gp$.

A QH lattice generalizes\,--\,in a sense made precise in Sects.\,2
and 3\,--\,the data of an odd, positive integer, $m$, characterizing
the celebrated Laughlin fluid~\cite{L} with Hall fraction
(dimensionless Hall conductivity) $\,\sH =1/m,\ m=1,3,5,\ldots\ $.
Localizable, finite-energy excitations above the ground state of a QH
fluid have ``charges'' corresponding to points in $\Gp$. The scalar
product of a point, $\q$, in $\Gp$ with the charge vector $\Q$ is the
electric charge of the excitation described by $\q$; the scalar
product of $\q$ with itself determines its statistical phase. Since
the scalar product of any pair of vectors in $\Gs$ is a {\em
rational\,} number, and since $\Gp$ is contained in (or equal to)
$\Gs$, statistical phases are rational multiples of $\pi$, and
localizable, finite-energy excitations of an (incompressible) QH
fluid thus exhibit rational fractional (anyon) statistics. QH
lattices thus encode fundamental quantum numbers of finite-energy
excitations of (incompressible) QH fluids.

The basic assumptions underlying our theoretical framework are
summarized in Sect.\,2, and their explicit implementation in the
form of an effective theory describing conserved current densities of
QH fluids is reviewed in Sect.\,3\mini. The precise mathematical
definition of a QH lattice is given in Sect.\,2, where, moreover, the
important notion of a {\em (``primitive'') chiral QH lattice (CQHL)}
is introduced. CQHLs are the ``basic building blocks'' of QH lattices
and form the main objects of study in the present paper. Physically,
they correspond to QH fluids that are {\em either\mini} electron-rich
{\em or\mini} hole-rich. Furthermore, finite, but macroscopic QH
samples classified by {\em chiral\,} QH lattices exhibit conserved
{\em edge currents\mini} that circulate along their boundaries in only
{\em one\mini} chiral sense.

In order to efficiently organize the classification of CQHLs,
$(\G,\Q)$, it is convenient to characterize them in terms of {\em
numerical invariants}. Such invariants are introduced in Sects.\,2
and~4. Among them, the following ones play a key role:

(i) the Hall fraction (or dimensionless Hall conductivity),
$\sH(\G,\Q)$, which is given by the squared length of the charge
vector $\Q$ and thus turns out to be a {\em rational\,} number;

(ii) the dimension, $N$, of the integral lattice $\G$, which equals
the number of independent, conserved current densities of the
corresponding QH fluid;

(iii) the discriminant, $\D(\G,\Q)$, of $\G$, i.e., the order of
the abelian group formed by the (equivalence) classes of elements
in $\Gs$ modulo $\G$, which is related to the denominator,
$\dH(\G,\Q)$, of the Hall fraction $\sH(\G,\Q)$; and

(iv) an invariant, denoted $\,\lMg$, that, physically, corresponds
to the smallest relative angular momentum of a certain pair of
identical excitations (with the quantum numbers of the electron);
(for the Laughlin fluid with $\,\sH=1/m,\ \lMg=m$).

For CQHLs, the invariants $\mini\lM$ and $\mini\sH\mini$ are related
by

\be
\lMg \:\geq\: \sH^{-1}(\G,\Q)\ ,
\label{lMgeq}
\ee

\noi
which is a simple consequence of the Cauchy-Schwarz inequality; see
Sect.\,4\mini.

In terms of the two invariants $N$ and $\,\lM$, we can formulate
a phenomenological {\em stability principle} which we shall appeal to
in our comparison between theory and experiment:

\proclaim\indent Stability Principle. A QH fluid described (in the
scaling limit) by a CQHL $(\G,\Q)$ is the\, {\em more stable}, the\,
{\em smaller\mini} the value of the invariant $\,\lMg$ and, given
the value of $\,\lMg$, the\, {\em smaller\mini} the dimension $N$
of $\,\G$.

\noi
(A measure for the stability of a QH fluid is, e.g., the width of the
plateau of $\mini\sH$, as a function of the filling factor,
corresponding to that QH fluid.) A detailed discussion of this
stability principle is contained in~\cite{FST}. Here it serves as a
``working hypotheses''. A consequence of our stability principle is
that QH fluids at values of $\sH$ that have {\em large
denominators\mini} are {\em unstable}.

If we are given physically plausible upper bounds, $\Ns$ and $\Ls$,
on the invariants $N$ and $\,\lM$ (see Sect.\,4) then the set of
CQHLs satisfying these bounds and thus physically observable can
be shown~\cite{FST} to be {\em finite}. In other words, in {\em
any\mini} interval of Hall fractions $\sH$, there are infinitely
many fractions that {\em cannot\,} be realized by a {\em
physically observable\mini} chiral QH fluid!

Inequality~\cfr{lMgeq} leads to a natural decomposition of the
interval $\,(0\mini,1]\,$ of Hall fractions $\mini\sH\mini$ into
subintervals (or ``{\em windows\,}''), $\Sigp$, with $\,1/(2\mini
p+1) \leq \sH < 1/(2\mini p-1)$. These windows can be further
divided into two halfs, $\,\Sigp=\Sigp^+\cup\Sigp^-$, where

\be
\Sigp^+ \::=\: [\mini{1\over 2\mini p+1}\mini,\mini {1\over 2\mini
p}\mini)\ , \hah
\Sigp^- \::=\: [\mini{1\over 2\mini p}\mini,\mini {1\over 2\mini
p-1}\mini)\ ,\hsp p=1,2,\ldots\ .
\label{sigpm}
\ee

By inequality~\cfr{lMgeq}, CQHLs with $\,\sH\in\!\Sp\,$ must satisfy
$\,\lM\geq 2\mini p+1$, and hence, by our stability principle, the
stability of QH fluids with $\,\sH\in\!\Sp\,$ {\em decreases}, as
$\mini p\mini$ {\em increases}.

Before summarizing further theoretical results, we pause to reflect
on the experimental data on Hall fractions in the interval $\,0<\sH
\leq 1$, that have been established in the literature on {\em
single-layer/component\,} QH systems; see Refs.~\cite{Tsui}
through~\cite{Eng}. These data are displayed in Fig.\,1. Indications
on the experimental status of the fractions are provided. In
Fig.\,1, we write $\,\sH=\nH/\dH\,$ and display the data in a
``$\dH$ versus $\mini\sH\mini$ plot''. Moreover, a grid delimiting
the subwindows $\Sp^\pm$ (see~\cfr{sigpm}) is superimposed on the
figure. Note that, in the interval of Hall fractions $\,0<\sH\leq
1$, there are {\em no\mini} experimental data {}from {\em
single-layer/component\,} QH systems showing the characteristics of
the QH effect at even-denominator fractions. Celebrated observations
on even-denominator QH fluids have been reported
in~\cite{evenSingle}, where a single-layer QH fluid with
$\,\sH=5/2\,$ is described, and in~\cite{evenDouble}, where
two-layer/component QH fluids with $\,\sH=1/2\,$ have been
established.


\addtocounter{page}{1}
Partly motivated by the experimental data on single-layer
QH systems, as collected in Fig.\,1, partly for theoretical reasons
(Wigner lattice instability), we expect a realistic upper bound on
$\,\lM$ to be given by

\be
\lM \:\leq\: \Ls \:=\: 7\;\,(\mbox{or }\,9)\ .
\ee

\noi
Assuming the bound $\,\Ls=7$, we predict that there are {\em no\mini}
(incompressible) chiral QH fluids with $\,\sH\in\!\Sp$, for
$\,p\geq 4$. Furthermore, we shall see that, in the subwindow
$\,\Sigma_3^+=[1/7,1/6)$, the {\em only\mini} physically realizable
Hall fractions are the elements of the series $\,\sH=N/(6N+1),\
N=1,2,\ldots\,$, and each such fraction is realized by a {\em
unique\mini} CQHL (of dimension $N$).

The situation in the ``complementary'' subwindow $\,\Sigma_3^-=
[1/6,1/5)\,$ is much more involved. It is a general consequence of
our analysis that, {}from a ``structural'' point of view, the
classification problems for QH lattices with Hall fractions $\sH$ in
the two complementary subwindows $\Sp^+$ and $\Sp^-,\
p=1,2,\ldots\,$, are {\em strikingly different\mini}. This fact is
reflected by the experimental data and will be illustrated by a
``uniqueness theorem'' proven in Sect.\,4\mini:

According to our stability principle, CQHLs $(\G,\Q)$ with a small
value of the invariant $\,\lMg$ (and not too high dimension $N$)
describe stable QH fluids. If the Hall fraction $\sH(\G,\Q)$ belongs
to ${\Sp}$ then, by~\cfr{lMgeq}, the minimal value of the invariant
$\,\lMg$ is given by $\,2\mini p+1,\ p=1,2,\ldots\,$, and
(``primitive'') CQHLs realizing this value are called {\em
\Lmini\mini}; see Sect.\,4. We shall see that all \Lmini\ CQHLs with
$\,\sH\in\!\Sp^+\,$ can be enumerated explicitly. There is a {\em
unique} $N$-dimensional CQHL with Hall fraction $\,\sH=N/(2\mini
pN+1),\ N,p=1,2,\ldots\,$, and the the edge states of the
corresponding QH fluids carry a representation of the Kac-Moody
algebra \suhN\ at level $1$; (for information on Kac-Moody
algebras, see~\cite{GO}). For $\,p\!\leq\!3\,$ and sufficiently small
values of the dimension $N$ (stability principle\mini!), the above
fractions correspond to experimentally well verified plateaux of
$\mini\sH$. It is interesting to note that, e.g., in
$\,\Sigma_1^+=[1/3,1/2)$, there are {\em no\mini} \Lmini\ CQHLs at
the fractions $\,\sH=4/11$, and $5/13\mini$; see Fig.\,1 and
Sect.\,5 for implications of this fact.

The classification of QH lattices (with Hall fractions) in the
subwindows, $\Sigma_p^-,\ p=1,2,\ldots\,$, is greatly facilitated by
the existence of a family of maps $\,\Sm,\ p=1,2,\ldots\,$, called
``{\em shift maps\,}''. These maps relate CQHLs of equal dimension
at shifted values of the {\em inverse\mini} Hall fraction,
$\,\Sm:\sH^{-1}\mapsto\sH^{-1}+2\mini p\,$; see Sect.\,4\mini.
Restricting their action to the class of \Lmini\ CQHLs, we shall see
that they yield one-to-one correspondences between the sets of
such lattices in the windows $\Sigma_1$ and $\Sigma_{p+1}$. Hence,
for the classification of \Lmini\ CQHLs in the subwindows $\Sp^-$, it
suffices to classify all \Lmini\ CQHLs in the ``fundamental
domain'' $\,\Sigma_1^-=[1/2,1)$, where $\,\lM=3\mini$. However, in
contrast, to the possibility of completely enumerating all
\Lmini\ CQHLs in the complementary fundamental domain $\,\Sigma_1^+=
[1/3,1/2)$, the classification of \Lmini\ CQHLs in $\Sigma_1^-$
is much more involved. In fact, this classification has been one of
the main objectives in our recent work~\cite{FST} where all
low-dimensional $(N\leq 4)$ ``indecomposable'', \Lmini\ CQHLs in
$\Sigma_1^-$ have been classified, as well as all those \Lmini\ CQHLs
in arbitrary dimension that exhibit ``large symmetries'' and thus
are called ``maximally symmetric''. (These latter CQHLs are natural
generalizations of the $su(N)$-QH lattices in $\Sigma_1^+$ mentioned
above.) The most relevant results of this classification are
recapitulated in Sect.\,5.  Here, we only mention one striking
feature thereof:

In general, one finds {\em more than one\mini} \Lmini\ CQHL realizing
a given Hall fraction $\sH\in\!\Sp^-,\ p=1,2,\ldots\,$, and the
different CQHLs at a given fraction $\sH$ typically form interesting
patterns of ``{\em QH lattice embeddings\,}''. Physically, these
embeddings find a natural interpretation in terms of possible {\em
phase transitions between ``structurally distinct'' QH
fluids\mini}. A theory of such phase transitions, which follow a
``symmetry breaking'' logic, has been developed
in~\cite[Sect.\,7]{FST}. The most likely Hall fractions
$\mini\sH\mini$ at which such structural phase transitions may
occur are found to be $\,2/3,\ 3/5,\ 4/7,\ 5/7,\ 5/9$, and
$\,1/2\mini$; compare with Fig.\,1$\mini$!

The main attention of the present paper is on {\em \Lmini, chiral\,}
QH lattices which we expect (stability principle) to describe the
most stable physical QH fluids. In Sect.\,5, we analyze to which
extent experimental data actually support the physical relevance of
the two concepts of \Lminiy\ and chirality. Explicit Hall fractions
are listed where new experimental data could lead to new theoretical
insights. Readers only interested in experimental consequences of our
analysis are invited to jump, after a short look at Sect.\,2,
directly to Sect.\,5, where our theoretical findings are summarized,
and their main phenomenological implications are discussed.


\begin{flushleft}
\section{QH Fluids and QH lattices: Basic Concepts}
\end{flushleft}

Our analysis of the fractional quantum Hall effect is based on a
precise notion of {\em incompressible (dissipation-free) quantum Hall
fluids} (or, for short, {\em QH
fluids\mini})~\cite{FSC,FSR,FT}\,--\,see assumptions {\bf (A1)}
through {\bf (A4)} below\,--\,from which  their main features can be
derived. Our mathematical characterization, in terms of ``QH
lattices'', of (universality classes of) QH fluids enables us to
enumerate and classify QH fluids. In this section, we review the
defining properties of QH lattices.

The fractional QH effect is observed in two-dimensional gases of
electrons at temperatures $\mini T \approx 0\,K\mini$ subject to a
very nearly constant magnetic field, ${\bf B}_{c}$, transversal to
the plane of the system. Two-dimensional gases of electrons can be
realized as heterojunctures~\cite{QHE}. Let $E_{0}$ denote the
ground-state energy of the system in a fixed magnetic field ${\bf
B}_{c}$. If there is a mobility gap $\mini\delta\mini$ {\em strictly
positive}, {\em uniformly\mini} in the size of the system, i.e., if
there are no {\em extended\,} (conducting) states in the spectral
subspace corresponding to the energy interval $\,[ E_{0} , E_{0} +
\delta ]\,$ then we say that the system is a QH fluid. More
precisely, a QH fluid is characterized by the property that
connected Green functions of the electric charge - and current
densities have cluster decomposition properties stronger than those
encountered in a system where the electric charge - and current
density couple the ground state of the system to a Goldstone boson
(a London superconductor). One can then show~\cite{FSC} that the
longitudinal resistance, $\RL$, of a QH fluid {\em vanishes}. This
can also be used as a {\em definition} of an (incompressible) QH
fluid. Let

\be
\nu \::=\: \frac{n}{(e B_{\!c}^\perp / h)}
\ee

\noi
be the filling factor of the system, where $\mini n\mini$ denotes
the electron density, $e\mini$ the elementary electric charge,
$B_{\!c}^\perp$ is the component of the constant magnetic field ${\bf
B}_{c}$ that is perpendicular to the plane of the system, and
$\mini h\mini$ is Planck's constant. The longitudinal resistance
$\RL$ is a complicated function of $\nu$, and it is a difficult
problem of many-body theory to predict where $\RL$ vanishes as a
function of $\nu$; see~\cite{QHE}. We do not solve this problem in
this paper. Instead, we show that {\em if\,} $\RL$ {\em
vanishes\mini} then the {\em Hall conductivity},

\be
\sH \:=\: \RH^{-1}\ ,
\ee

\noi
necessarily belongs to a certain set of {\em rational multiples} of
$e^2/h$, and, given such a value of $\sH$, we can determine the
possible types of quasi-particles, i.e., the different ``Laughlin
vortices'' of the system, their electric charges and their
statistical phases; see~\cite{WB,FZ,S,FSR,FT}.

Next, we describe the basic assumptions and physical principles underlying our
analysis of QH fluids.
\rule[-4mm]{0mm}{5mm}

{\bf (A1)} The temperature $T$ of the system is close to $0\,K$. For
an (incompressible) QH fluid at $T=0\,K$, the {\em total electric
charge} is a good quantum number to label physical states of the
system describing excitations above the ground state;
see~\cite{FT,FGM}. The charge of the ground state of the system is
normalized to be zero.
\rule[-4mm]{0mm}{5mm}

{\bf (A2)} In the regime of very short wave vectors and low
frequencies, the {\em scaling limit}, the total electric current
density is the sum of $\,N=1,2,3,\ldots$ {\em separately conserved\,}
\ui-current densities, describing electron and/or hole transport in
$N$ separate ``channels'' distinguished by conserved quantum
numbers. (For a finite, but macroscopic sample, this assumption
implies that there are $N$ separately conserved chiral \ui-edge
currents~\cite{H} circulating round the boundary of the system.) In
our analysis, we regard $N$ as a free parameter.\,--\,Physically,
$N$ turns out to depend on the filling factor $\nu$ and other
parameters characterizing the system.
\rule[-4mm]{0mm}{5mm}

{\bf (A3)} In our units where $\,h=-e=1$, the electric charge of an
{\em electron/hole} is $+1/-1$. Any local excitation
(quasi-particle) above the ground state of the system with {\em
integer\mini} total electric charge $q_{el}$ satisfies {\em
Fermi-Dirac statistics\mini} if  $q_{el}$ is {\em odd}, and {\em
Bose-Einstein statistics\mini} if $q_{el}$ is {\em even}.
\rule[-4mm]{0mm}{5mm}

{\bf (A4)} The quantum-mechanical state vector describing an {\em
arbitrary\mini} physical state of an (incompressible) QH fluid is
{\em single-valued\,} in the position coordinates of all those
(local) excitations that are composed of {\em electrons\mini}
and/or {\em holes}.
\rule[-4mm]{0mm}{5mm}

The basic contention advanced in ~\cite{FSC,FSR,FT} is that if a
QH fluid is interpreted as a two-dimensional system of electrons
with vanishing longitudinal resistance $\RL$, satisfying
assumptions {\bf (A1)} through  {\bf (A4)} above, then, in the
scaling limit, its quantum-mechanical description is completely
coded into a {\em quantum Hall lattice (QH lattice)}. A QH lattice
$(\G,\Q)$ consists of an odd, integral lattice $\G$ and an
integer-valued, linear functional $\Q$ on $\G$. In general, the
metric on $\G$ need not be positive definite. The number of positive
eigenvalues of the metric corresponds, physically, to the number of
edge currents of one chirality, the number of negative eigenvalues
correponds to the number of egde currents of opposite chirality. In
this paper, we present results on QH fluids with edge currents of
just {\em one\mini} chirality. These QH fluids correspond to QH
lattices $(\G,\Q)$ where $\G$ is a {\em euclidian\mini} lattice,
i.e., a lattice with a {\em positive-definite\mini} metric. These
special QH lattices are called {\em chiral QH lattices (CQHLs)}.

Before explaining more precisely what a CQHL is, we note that a
physical hypothesis expressing a ``{\em chiral factorization\,}''
property of QH fluids motivates our study of {\em chiral\,} QH
lattices.
\rule[-4mm]{0mm}{5mm}

{\bf (A5)} The fundamental charge carriers of a QH fluid are
electrons and/or holes. We assume that, in the scaling limit, the
dynamics of electron-rich subfluids of a QH fluid is {\em
independent\mini} of the dynamics of hole-rich subfluids, and, up to
charge conjugation, the theoretical analysis of an electron-rich
subfluid is {\em identical\,} to that of a hole-rich subfluid.
\rule[-4mm]{0mm}{5mm}

A discussion of the ``working hypothesis''~{\bfÊ(A5)}, including
some proposals for its experimental testing, is given in~\cite{FST}.
There it is also shown that all ``hierarchy fluids'' of the
Haldane-Halperin~\cite{HH} and Jain-Goldman~\cite{JG} scheme,
respectively, satisfy our assumptions {\bf (A1--4)}, and the status
of assumption {\bf (A5)} is carefully analyzed for such fluids;
(see especially Appendix~E in~\cite{FST}).

\vspace{5mm}

\noi
{\bf Chiral Quantum Hall Lattices.} Let $V$ be an $N$-dimensional
real vector space equipped with a {\em positive-definite} inner
product, $\la.\,,.\ra\!$. In $V$ we choose a basis $\{ {\bf
e}_{1},\ldots,{\bf e}_{N}  \}$ such that

\be
K_{ij} \:=\: K_{ji} \::=\:\: \sca{{\bf e}_{i}\mini}{{\bf e}_{j}}
\,\in \Z\ , \hfa i,j = 1, \ldots,N\ ,
\label{K}
\ee

\noi
i.e., {\em all\mini} matrix elements are {\em integers}. The basis
$\{ \avecd{{\bf e}} \}$ generates an {\em integral, euclidean} (i.e.,
positive definite) {\em lattice}, $\G$, defined by

\be
\G \::=\: \{\, \q = \sum_{i=1}^{N} q^i \e_i \,|\  q^i \in \Z\mini,
\mbox{ for all } i=1,\ldots,N \, \}\ .
\label{G}
\ee

\noi
The matrix $K$ in~\cfr{K} is called the {\em Gram matrix\mini} of
the basis $\{ \avecd{{\bf e}} \}$ generating the lattice $\G$. Let
$\{ \avecu{\es} \}$ be the basis of $V$ {\em dual\,} to $\{
\avecd{\e} \}$, i.e., the basis satisfying

\be
\sca{\es^{i}}{\e_{j}} \:=\: \delta_j^i\ , \hfa i,j = 1,\ldots,N
\ .
\ee

\noi
Then,

\be
\es^i \:=\: \sum_{j=1}^N ( K^{-1} )^{ij} \e_j\ ,
\label{epsi}
\ee

\noi
where $K^{-1}$ is the inverse of the matrix $K = ( K_{ij} )$ with $K_{ij}$
given in~\cfr{K}, and $( K^{-1} )^{ij} =\; \sca{\es^{i}}{\es^{j}}$.
The basis $\{ \avecu{\es} \}$ generates the {\em dual lattice},
$\Gs$, defined by

\bea
\Gs & := & \{ \, \n \in V \,|\ \sca{\n\,}{\q} \in \Z\mini, \mbox{ for
all } \q \in\!\G \,\}  \nonumber \\
& = & \{\, \n = \sum_{i=1}^{N} n_i\, \es^i \,|\  n_i
\in \Z\mini, \mbox{ for all } i=1,\ldots,N \, \}\ ,
\label{Gs}
\eea

\noi
and, by~\cfr{epsi}, $\Gs$ contains $\G$. Denoting by $\,\D := \det K
\in\!\Z\,$ the determinant of the matrix $K$ ($\det K > 0\,$ for
euclidean lattices), the matrix elements $( K^{-1} )^{ij}$ giving the
scalar products between basis vectors in $\Gs$ are, in general (for
$\D \neq 1$), {\em rational\,} numbers; since, by Kramer's rule, $(
K^{-1} )^{ij}=( \widetilde{K} )^{ij}/\D$, with  $\widetilde{K}$
the integer-valued matrix of cofactors of the matrix $K$.

The (equivalence) classes of elements in $\Gs$ modulo $\G\mini$ form
an abelian group, $\Gs / \, \G$. There are $\,\D = | \Gs/\,\G|\,$
distinct classes. $\D$ is often referred to as the lattice
{\em discriminant}. An integral lattice $\G$ is said to be {\em
odd\,} if it contains a vector $\q$ such that $\,\sca{\q\,}{\q}$ is
an {\em odd\,} integer. Thus $\G$ is odd if and only if $K_{ii} =\:
\sca{\e_{i}\mini}{\e_{i}}$ is odd for at least one $i$. (Otherwise,
$\G$ is said to be {\em even}.)

Given a set of integers, $\avecd{n}$, we denote by
$\,\gcd(\avecd{n})\,$ their {\em greatest common divisor}. A vector
$\,\n = \sum_{i=1}^N n_i\, \es^i \in\!\Gs\,$ is called {\em
primitive} (or ``visible'') if and only if

\be \gcd(\avecd{n}) \:=\: \gcd (
\sca{\n\,}{\e_1},\ldots,\sca{\n\,}{\e_N} ) \:=\: 1\ .
\label{primi}
\ee

\noi
In geometrical terms, a vector $\,\n \in\!\Gs\,$ is primitive if and
only if the line segment joining the origin to the point $\,\n\,$ is
free of any lattice point. In particular, one can always take a
primitive vector as the first vector of a lattice basis.

A lattice $\G$ is said to be {\em decomposable\mini} if it can be
written as an orthogonal direct sum of sublattices,

\be
\G \:=\: \bigoplus_{j=1}^r \Gamma_j \ , \hfs r \geq 2 \ ,
\label{Gd}
\ee

\noi
with the property that $\,\sca{\q^{(i)}}{\q^{(j)}}\:= 0$, for
arbitrary vectors  $\q^{(i)}$ in $\Gamma_i$  and  $\q^{(j)}$ in
$\Gamma_j$, and for arbitrary $i \neq j$. Otherwise, $\G$ is said to
be {\em indecomposable}.

If $\,m\,$ and $\,n\,$ are two integers we shall write $\,m \equiv n
\bmod p\,$ if and only if $\,m-n\,$ is an integer multiple of $p$.

We are now prepared to define what we mean by a {\em chiral
QH lattice (CQHL)}:

\proclaim\indent Definition. A CQHL is a pair $(\G,\Q)$ where $\G$ is
an odd, integral, euclidean lattice and $\Q$ is a primitive vector in
the dual lattice $\,\Gs$ with the property that
\be
\sca{\Q\,}{\q} \:\equiv\: \sca{\q\,}{\q}\:\bmod\ 2\ ,\hfa\,\q \in\!
\G\ .
\label{parity}
\ee

Let $( \G , \Q )$ be a CQHL for which $\,\G = \bigoplus_{j=1}^r
\Gamma_j \,$ is  decomposable. Then $\,\Gs = \bigoplus_{j=1}^r
\Gamma_j^\ast \,$ is the associated decomposition of the dual
lattice. We say that $(\G,\Q)$ is {\em proper\mini} if, in the
decomposition of $\Q$

\be
\Q \:=\: \sum_{j=1}^r \Q^{(j)}\ , \hwh \Q^{(j)} \::=\:
\Q\!\mid_{\Gamma_j^\ast}\:\in\!\Gamma_j^\ast \ ,
\label{Qd}
\ee

\noi
corresponding to the decomposition of $\,\Gs$, every $\mini\Q^{(j)}$
is {\em non-zero}.

Finally, we introduce the following notion which plays an important
role in the sequel:

\proclaim\indent Definition. A CQHL is called\, {\em primitive} if,
in the decomposition~\cfr{Qd} of $\,\Q$, the component $\,\Q^{(j)}$
is a\, {\em primitive} vector in $\Gamma_j^\ast$, for all $\,j =
1,\ldots, r$.

We consider this primitivity property to be a natural requirement on
{\em composite} CQHLs that correspond to physically observable
QH fluids; see the discussion at the end of Sect.\,3\mini. We note
that any {\em indecomposable} CQHL, i.e., one with $r=1$
in~\cfrs{Gd}{Qd}, is proper and primitive.


\begin{flushleft}
\section{A Dictionary Between the Physics of QH Fluids and the
Mathematics of QH Lattices}
\end{flushleft}

In this section, we construct a precise dictionary between
mathematical properties of QH lattices and physical properties of QH
fluids. Such a  dictionary has already been presented
in~\cite{FSC,FSR,FT}. Here we just recall its main contents and
significance. The starting point is the idea to describe the physics
of a QH fluid in the scaling limit in terms of an effective field
theory of its conserved current densities. Since a QH fluid has a
strictly positive mobility gap $\delta$, the scaling limit of the
effective theory of its conserved current densities must be a {\em
topological field theory}~\cite{FZ,FSC}. The presence of a non-zero
external magnetic field transversal to the plane to which the
electrons of a QH fluid are confined implies that the quantum
dynamics of the system {\em violates} the symmetries of parity
(reflection-in-lines) and time reversal. Thus the topological field
theory will {\em not\,} be parity- and time-reversal invariant.

In $2 + 1$ space-time dimensions, the continuity equation

\be
{\partial\over \partial t}\, j^0 + \vec{\bigtriangledown}\cdot\vec{j}
\:=\: 0\ , \ee

\noi
obeyed by a conserved current density $\,j^\mu=(j^0,\vec{j}\,)$,
implies that $j^{\mu}$ is the curl of a vector potential, i.e.,
$j^{\mu} = \varepsilon^{\mu \nu \lambda} \, \partial_{\nu}
b_{\lambda}$, for a vector potential $\,b_{\lambda} =
(b_0,\vec{b}\,)\,$ which is unique up to the gradient of a scalar
field (gauge invariance!).

Let us consider a QH fluid which, in the scaling limit, has $N$
independent, conserved current densities $\,j^{\mu\mini
1},\ldots,j^{\mu\mini N}$, the electric current density,
$J_{el}^\mu$, being a linear combination, $J_{el}^\mu=\sum_{i=1}^N
Q_i\, j^{\mu\mini i}$, of these current densities. When formulated in
terms of the vector potentials $\,\avecu{b_{\lambda}}\,$ of these
conserved current densities, the topological field theory describing
the physics of the {QH fluid} in the scaling limit can only be a {\em
pure, abelian Chern-Simons theory\mini} in the fields
$\,\avecu{b_{\lambda}}$, as follows {}from the circumstance that it
must violate parity - and time-reversal invariance. This has been
shown in refs.~\cite{FK,FSC,FSR,FT}. The bulk action is given by

\be
S( {\bf b}) \:=\: \frac{1}{4 \pi}
\int C_{ij} \, b_{\mu}^i\, \partial_{\nu} b_{\lambda}^j \,
\varepsilon^{\mu \nu \lambda}\,dt\diix \:=:\: \frac{1}{4 \pi} \int
\sca{{\bf b}_{\mu}\,}{ \partial_{\nu} {\bf b}_{\lambda}}
\!\varepsilon^{\mu \nu \lambda}\,dt\diix\ ,
\ee

\noi
where $\,C_{ij}= C_{ji}\,$ is some non-degenerate quadratic form
(metric) on \BBn{R}{N}.

Physical states of a pure, abelian Chern-Simons theory describe
static $N$-tuples of ``charges'' localized in bounded disks of
space. Each $N$-tuple, $(\avecu{q})$, of ``charges'' localized in
some disk $D$ of space is an $N$-tuple of eigenvalues of the
operators

\be
\int\limits_D j^{0i} (\vec{x},t)\,\diix \:=\:
\oint\limits_{\partial D}  \vec{b}^{\mini i} (\vec{x},t)\cdot
d\vec{x} \ ,\hsp i = 1,\ldots,N\ ,
\ee

\noi
acting on a corresponding physical state of the theory. The equations
of motion of pure, abelian Chern-Simons theory, with the currents
$j^{\mu i}$  minimally coupled to $N$ external gauge fields $a_{\mu
i}$, $i=1,\ldots,N$, read

\be
\varepsilon_{\mu \nu \rho} j^{\rho\mini i} \:=\: \partial_{\mu}
b_{\nu}^i - \partial_{\nu} b_{\mu}^i \:=\:  (C^{-1})^{ij} \left(
\partial_{\mu} a_{\nu j} - \partial_{\nu} a_{\mu j} \right)
\:=:\:  (C^{-1})^{ij} f_{\mu \nu j} \ ,
\label{CSeq}
\ee

\noi
These equations imply that, to an $N$-tuple of ``charges''
$(\avecu{q})$, there corresponds an $N$-tuple of ``fluxes'',
$(\avecd{\varphi})$, with $\,\varphi_j = \int_D f_{12j}\,\diix\,$
related to $q^i$ through

\be
q^i \:=\: (C^{-1})^{ij} \varphi_j\ ,\hsp  i=1,\ldots,N\ .
\label{qphi}
\ee

Consider a state of Chern-Simons theory describing two excitations
with {\em identical\,} ``charges'' $(\avecu{q})$ localized in
disjoint, congruent disks of space, $D_1$ and $D_2$. {}From the
theory of the {\em Aharonov-Bohm effect}~\cite{AB} we know that if
the positions of the two disks are exchanged adiabatically along
counter-clockwise oriented paths, the state vector only changes by
a phase factor, $\exp i\pi\theta$, given by

\be
\exp {i\pi\theta} \:=\: \exp{i\pi\sca{\q\,}{\q}} \ ,
\ee

\vwv

\be
\theta \:=\: \sum_{i=1}^N \varphi_i \, q^i
\:=\: \sum_{i=1}^N (C^{-1})^{ij} \varphi_i\, \varphi_j
\:=\: \sum_{i=1}^N C_{ij}\, q^i q^j
\:=:\: \sca{\q\,}{\q} \ .
\ee

\noi
It is well known~\cite{StatP} that $\,\exp i\pi\!\sca{\q\,}{\q}$
has the meaning of a {\em statistical phase} of the excitation
corresponding to the ``charges'' $(\avecu{q})$.  The
$N$-dimensional vector $\q\,$ introduced here is equivalently
defined through its components $(\avecu{q})$, which are ``charges'',
or, through~\cfr{qphi}, in terms of its dual (w.r.t.~the metric
$C_{ij}$) components $(\avecd{\varphi})$, the ``fluxes''. An
excitation $\q\,$ with $\,\mbox{$\sca{\q\,}{\q}$} \equiv 1 \bmod
2\,$ is a fermion, while if  $\,\sca{\q\,}{\q} \equiv 0 \bmod 2\,$ it
represents a boson.

Next, we consider a state describing two excitations with two
{\em different\mini} charge vectors ${\bf q}_{(1)}$ and ${\bf
q}_{(2)}$ localized in disjoint disks $D_1$ and $D_2$. Imagine that
$D_2$ is transported around $D_1$ adiabatically, along a
counter-clockwise oriented loop. Then the theory of the
Aharonov-Bohm effect teaches us that the state vector only changes
by a phase factor given by

\be
\exp {2 i\pi\sca{{\bf q}_{(1)}\mini}{{\bf q}_{(2)}}}\ ,
\label{monodr}
\ee

\vwv

\be
\sca{{\bf q}_{(1)}\mini}{{\bf q}_{(2)}}
\::=\: \sum_{i=1}^N C_{ij}\, q_{(1)}^i q_{(2)}^{\mini j}
\:=\:\sum_{i=1}^N \varphi_{(1)i} \, q_{(2)}^i
\:=\: \sum_{i=1}^N (C^{-1})^{ij} \varphi_{(1)i}\,\varphi_{(2)j}
\ee

Since the ``charges'', $q^i$, of physical states\,--\,i.e., the
eigenvalues of the operators $\int_D j^{0\mini i}
(\vec{x},t)\,\diix,\ i=1,\ldots,N$, on physical states\,--\,are
{\em additive\mini} quantum numbers, the set of vectors
$\mini\q\mini$ of physical excitations of a QH fluid form a {\em
lattice}, $\Gp$, whose dimension can be taken to be $N$. Expressing
the electric current density $J_{el}^\mu$ as a linear combination,
$J_{el}^\mu=\sum_{i=1}^N Q_i\,j^{\mu\mini i}$, of the current
densities $\,j^{\mu\mini 1},\ldots,j^{\mu\mini N}$, the {\em total
electric charge} (in units of $-e$), $q_{el}(\q)$, of an excitation
of the system with charge vector $\q$ localized in some disk $D$ is
found to be given by

\be
q_{el}(\q) \:=\: \sum_{i=1}^N Q_i\,q^i \:=:\; \sca{\Q\,}{\q}\ ,
\ee

\noi
where the electric charge assignments $Q_i$ of each current have
been collected in an $N$-dimensional vector $\Q$ henceforth
referred to as the ``charge vector''. Note that, according to the
above definition, the $Q_i$ are the {\em dual\,} components of the
charge vector $\Q$. The electric charge $q_{el}(\q)$ is an
eigenvalue of the electric charge operator

\be
\int\limits_D J_{el}^0 (\vec{x},t)\,\diix \:=\:
\int\limits_D \left( \sum_{i=1}^N Q_i \, j^{0\mini i}(\vec{x},t)
\right)\diix\ .
\ee

We define $\G$ as the set of all physical excitations $\q$ with {\em
integral\,} electric charge, i.e.,

\be
\G \::=\: \{\, \q \in \Gp \,|\ \sca{\Q\,}{\q}=q_{el}(\q) \in \Z
\,\}\ .
\ee

\noi
Clearly $\G$ is a sublattice of $\,\Gp$. If $\,\G$ has at least one
excitation $\q$ of electric charge $\,q_{el}(\q) = 1\,$ then the
dimension of $\,\G$ is equal to $N$. Assumption~{\bf (A3)} entails
that arbitrary vectors $\q_{(i)}$ in $\G$ must  have integer
statistical phases $\,\sca{{\bf q}_{(i)}\mini}{{\bf q}_{(i)}}$
related to their electric charges by the congruence $\,q_{el}({\bf
q}_{(i)}) \equiv\:\sca{{\bf q}_{(i)}}{{\bf q}_{(i)}} \bmod\
2\mini$. Writing the scalar product $\,\sca{{\bf
q}_{(1)}\mini}{{\bf q}_{(2)}}$ between two arbitrary charge
vectors in $\G$ as

\ben
\sca{{\bf q}_{(1)}\mini}{{\bf q}_{(2)}} \:=\: \frac{1}{2}\left(
\sca{{\bf q}_{(1)}+{\bf q}_{(2)}\mini}{{\bf q}_{(1)}+{\bf q}_{(2)}}
- \sca{{\bf q}_{(1)}\mini}{{\bf q}_{(1)}} - \sca{{\bf
q}_{(2)}\mini}{{\bf q}_{(2)}} \right) \ ,
\een

\noi
we see that this is always an  integer and  conclude that $\G$ is
an {\em integral\,} lattice. Since there is at least one
electron-like excitation $\q$ in $\G$ with $\,q_{el}(\q)=1$, {\bf
(A3)} also constrains the lattice $\G$ to be {\em odd\,} because
$\,\sca{\q\,}{\q}  \equiv q_{el}(\q) \equiv 1 \bmod 2\mini$. For
an arbitrary lattice basis $\{ \q_{(1)},\ldots,\q_{(N)}
\}$ of $\,\G$, all the scalar products $\,Q_i=\:
\sca{\Q\,}{\q_{(i)}},\ i=1,\ldots,N$, are integers, and hence the
vector $\Q$ belongs to the {\em dual}, $\Gs$, of $\,\G$
(see~\cfr{Gs}). The charge vector $\Q$ must necessarily also be {\em
primitive}\,--\,i.e., $\gcd(\avecd{Q}\, )=1$\,--\,in order for an
excitation of electric charge $1$ to exist.

Every vector $\q$ of the sublattice $\G$ of $\,\Gp$ can now be
consistently interpreted as a physical state describing a
(multi-)electron and/or (multi-)hole configuration (depending on
the integral value of the electric charge) excited {}from the ground
state of the system. Note that, starting {}from an electron-like
excitation $\q_{(1)}$ in $\G$, one can form a full basis for $\G$
consisting of electron vectors, $\q_{(i)}$, with
$\,q_{el}(\q_{(i)})=1$, for $\,i=1,\ldots,N$. We call such a
basis a ``symmetric'' lattice basis.

The next step consists in finding restrictions on the lattice $\Gp$
by making use of our last assumption {\bf (A4)}. Consider, for each
$\,i=1,\ldots,N$, a physical state of the system describing an
excitation with ``charges'' $\q^\prime$ localized in a disk $D_1$
and an excitation corresponding to an electron with vector
$\q_{(i)}$ localized in a disk $D_2$ disjoint {}from $D_1$. Then we
derive {}from assumption {\bf (A4)} and~\cfr{monodr} that

\be
\exp 2\mini \pi i \sca{\q^\prime}{\q_{(i)}} \;=\: 1 \ ,
\ee

\noi
i.e., $\sca{\q^\prime}{\q_{(i)}}$ must be an {\em integer}, for
all $\,i=1,\ldots,N$. {}From this, it follows that $\q^\prime$
belongs to the dual lattice $\Gs$. Thus, vectors $\q^\prime$ of
{\em arbitrary\mini} physical excitations of a QH fluid described
by the lattice $\G$ and the charge vector $\Q$ must all belong to
the lattice $\Gs$ dual to $\G$. We then have the following
inclusions between the various lattices introduced so far:

\be
\Gs \:\supseteq\: \Gp \:\supseteq\: \G\ .
\ee

To discover how the data $(\G,\Q)$ predict the value of the Hall
conductivity $\sH$, we consider the effect of turning on a
perturbing external magnetic field $\,{\bf B}={\bf B}_{total} - {\bf
B}_{c}\,$ localized in a small disk $D$ inside the system and
creating a total magnetic flux $\Phi$, with

\be
\Phi \:=\: \int\limits_D B (\vec{x})\,\diix \ ,
\ee

\noi
where $B\,$ is the component of $\,{\bf B}$  perpendicular to the
plane of the system. Since the total electric current density
$J_{el}^\mu$ is given by $\,J_{el}^\mu=\sum_{i=1}^N
Q_i\,j^{\mu\mini i}$, of the current densities $\,j^{\mu\mini
1},\ldots,j^{\mu\mini N}$, the ``flux'' $\varphi_i$ created by
$\,{\bf B}$ is given by

\be
\varphi_i \:=\: Q_i \, \Phi \ ,
\label{fluxB}
\ee

\noi
for $\,i=1,\ldots,N$. This follows {}from the way in which, in
Chern-Simons theory, the currents, $j^{\mu \mini i}$, are coupled
to an external gauge field. Let $\,A_\mu = (A_0,\vec{A}\,)\,$ denote
the electromagnetic vector potential of the external magnetic field
$\,{\bf B}$. Then, {}from the fact that the total electric current
$J_{el}^\mu$ couples to the vector potential $A_\mu$ through the
term $\,J_{el}^\mu\,A_\mu$, it follows that the currents $j^{\mu\mini
i}$ are minimally coupled to the field $\,Q_i\mini A_\mu\mini$. Thus
the gauge fields $\,a_{\mu\mini i}\,$ appearing on the r.h.s.\
of~\cfr{CSeq} are given by $\,a_{\mu\mini i} = Q_i\mini A_\mu\mini$
which implies~\cfr{fluxB}; see~\cite{FSC,FSR,FT}. Since the charges
$\,q^i = \int_D  j^{0\mini i} (\vec{x},t)\,\diix = \oint_{\partial
D}\vec{b}^i (\vec{x},t)\cdot d\vec{x}$, corresponding to the fluxes
$\,\varphi_k\mini$, are equal to $\,(C^{-1})^{ik}\, \varphi_k\mini$,
by the equations of motion~\cfr{CSeq} of Chern-Simons theory, we
conclude that the {\em total excess electric charge} created by the
excess magnetic flux $\Phi\mini$ is given by

\bea
q_{el}(\Phi) &=& \sum_{i=1}^N
Q_i \, q^i \:=\: \sum_{i,k=1}^N Q_i \, (C^{-1})^{ik}\,\varphi_k
\nonumber \\
&=& \left( \sum_{i,k=1}^N Q_i \,(C^{-1})^{ik}\,Q_k \, \right) \Phi
\:=:\:\: \sca{\Q\,}{\Q}\!\Phi\ .
\label{chargePhi}
\eea

\noi
Comparing this equation to the equations describing the
electrodynamics of a QH fluid which have been described
in~\cite{FK,FSR}, in particular to the equation

\be
J_{el}^0 \:=\: \sH \, B \ ,
\ee

\noi
we find that the coefficient of $\mini\Phi$ on the
r.h.s.\ of~\cfr{chargePhi} is the {\em Hall conductivity}, $\sH$, of
the QH fluid. Thus, we arrive at the fundamental equation

\be
\sH \:=\:\: \sca{\Q\,}{\Q}\ .
\label{sH}
\ee

Since, as shown above, $\Q$ belongs to the dual lattice $\Gs$, we
have that

\ben
\Q \:=\: \sum_{i=1}^{N} Q_i \es^i
\een

\noi
with $Q_i\in\!\Z\mini$, for $\,i=1,\ldots,N$, where $\{ \avecu{\es}
\}$ is the basis of $\,\Gs$ dual to a basis $\{ \avecd{\e} \}$ of
the {\em integral\,} lattice $\G$, with Gram matirx $\,K_{ij}
=\:\sca{{\bf e}_{i}\mini}{{\bf e}_{j}} \in\!\Z\mini$. We could
choose, for example, a symmetric basis of electron-like excitations,
i.e., $\e_i =\q_{(i)}$, for all $\,i=1, \ldots,N$. In this
situation, the electric charge requirements fix the coefficients
$\,Q_i=1$, for all $\,i=1,\ldots,N$, since we must have
$\,q_{el}(\e_i)=\:\sca{\Q\,}{\e_i}=Q_i=1\mini$. But any other choice
of basis is admissible as well. The all important consequence of
$\mini\Q\mini$ being a vector in the dual of $\,\G$ is that its
squared length $\,\sca{\Q\,}{\Q}$ is a {\em rational number}. To see
this, we return to the discussion below~\cfr{Gs}. We have that
$\,\sca{\es^i}{\es^j}=( K^{-1} )^{ij}=( \widetilde{K} )^{ij}/\D$,
where $\,\D = \det K\,$ (the lattice discriminant) is an integer,
and $\,(\widetilde{K})^{ij}\,$ are integers. It follows that

\be
\sca{\Q\,}{\Q} \:=\; \sum_{i,j=1}^N Q_i\, Q_j\sca{\es^i}{\es^j}
\;=\: \left( \sum_{i,j=1}^N Q_i\,( \widetilde{K} )^{ij}\, Q_j
\right)\,\D^{-1}
\label{Qrat}
\ee

\noi
is a rational number whose denominator is a {\em divisor} of
$\D\mini$; (in general, there will be non-trivial common divisors of
$\D$ and of the numerator of the expression on the r.h.s.
of~\cfr{Qrat}). We conclude that the Hall conductivity $\sH$ of a QH
fluid satisfying assumptions {\bf (A1)} through {\bf (A4)} is
necessarily a {\em rational number}.

The analysis just completed shows that, in the scaling limit, the
physics of a QH fluid is described by a pair $(\G,\Q)$ of an
integral, odd lattice $\G$ and a primitive vector $\Q$ in the dual
lattice, with $\,\sca{\Q\,}{\q} \equiv\:\sca{\q\,}{\q} \bmod\ 2$,
for all vectors $\q$ in $\G$, i.e., by data that we have termed {\em
QH lattice\mini} in Sect.\,2\mini.

In the following, we specialize to {\em chiral\,} QH lattices
(CQHLs) which are QH lattices where the lattice $\G$ is
{\em euclidean}. They describe (universality classes of) QH fluids
with edge currents of {\em definite chirality}, the basic charge
carriers being, say, electrons. The theory of an electron-rich QH
fluid differs {}from that of a hole-rich QH fluid only in relative
minus signs in all equations relating charges to fluxes and
involving the electric charge of the basic charge carriers. One
simply reverses the sign of the charge vector $\Q$ and the metric
$C:\ \Q \rightarrow -\mini\Q\,$ and $\,C_{ij} \rightarrow -\mini
C_{ij}$,  i.e., $\sca{.\,}{.} \rightarrow -\sca{.\,}{.}\!$. A more
general type of QH fluids consisting of electron- {\em and\,}
hole-rich subsystems is therefore described by {\em two\mini} CQHLs,
$(\G_e,\Q_e)\, \mbox{ and } \, (\G_h,\Q_h)$, with

\be
\sH \:=\:\: \sca{\Q_e\mini}{\Q_e\mini} - \:\sca{\Q_h}{\Q_h}\ .
\ee

\noi
In such QH fluids the subsystems described by $(\G_e,\Q_e)$ and
by $(\G_h,\Q_h)$ are independent of each other; (in particular,
left- and right moving edge excitations are independent of each
other). The mathematical properties of $(\G_e,\Q_e)$ and of
$(\G_h,\Q_h)$ are analogous. It is therefore sufficient to study
the properties of $(\G_e,\Q_e)$, say, and we shall omit the
subscript ``$e$'' henceforth. Note that this ``factorized''
situation implements assumption~{\bf (A5)} of Sect.\,2\mini.

As in~\cfrs{Gd}{Qd}, let $\,\G = \bigoplus_{j=1}^r \Gamma_j \,$ be
the decomposition of the lattice $\G$ into an orthogonal direct sum
of indecomposable sublattices $\G_j$, and let $\,\Q = \sum_{j=1}^r
\Q^{(j)}$, with $\,\Q^{(j)} \in \Gamma_j^\ast$, be the associated
decomposition of the vector $\Q$ of electric charges. Then,
by~\cfr{sH},

\be
\sH \:=\:\: \sca{\Q\,}{\Q} \:=\: \sum_{j=1}^r
\sca{\Q^{(j)}}{\Q^{(j)}} \:=\: \sum_{j=1}^r \sH^{(j)}
\ee

\noi
is the corresponding decomposition of the Hall conductivity (or
{\em Hall fraction}) as a sum of Hall fractions of subfluids
described by the pairs $(\G_j,\Q^{(j)}),\ j=1,\ldots,r$. Let us
imagine that, for some $j$, $\Q^{(j)}=0$. Then $\sH^{(j)}=0$, and
the subfluid corresponding to $(\G_j,\Q^{(j)})$ does not have any
interesting electric properties. For the purpose of describing
electric properties of QH fluids and classifying the possible values
of the Hall fraction $\sH$ of QH fluids, subfluids described by
$(\G_j,\Q^{(j)})$, with $\Q^{(j)}=0$, can therefore be discarded.
Thus, we may assume henceforth that $\Q^{(j)}\neq 0$, for all
$\,j=1,\ldots,r$, i.e., we may limit our analysis to {\em
proper\mini} CQHLs; see~\cfr{Qd}. (However, if there are QH fluids
with spin-charge separation, subfluids $(\G_j,\Q^{(j)})$, with
$\Q^{(j)}=0$, will appear, and spin currents could receive
contributions {}from such neutral subfluids; see~\cite{FSR}.)

Next, suppose that, for some $j$ with $1\leq j \leq r$, $\,\Q^{(j)}$
is an integer multiple of a {\em primitive} vector $\,\Q_\ast^{(j)}
\in\!\G_j^\ast$, i.e., $\,\Q^{(j)}= \nu_j \, \Q_\ast^{(j)}$, with
$\nu_j \geq 2\mini$; see~\cfr{primi}. Then, for {\em any} $\,\q\in\!
\G_j,\ \,\sca{\Q^{(j)}}{\q\mini}\,$ is an {\em integer multiple} of
$\,\nu_j$. This would mean that the electric charge of an {\em
arbitrary\mini} quasi-particle of the subfluid described by
$(\G_j,\Q^{(j)})$ would be an integer multiple of $\nu_j$ (in units
where $-e=1$), i.e., only bound states of electrons and holes of
electric charge $\,n\mini\nu_j,\  n \in\!\Z\mini$, would appear as
quasi-particles of such a subfluid. There appear to exist QH fluids
where this situation arises (e.g. ``hierarchy QH fluids'', see
Appendix~E of~\cite{FST}, or films of superfluid $\mbox{He}^3$,
see~\cite{FSR}). For simplicity, we shall, however, assume henceforth
that $\,\nu_j=1$, i.e., that $\Q^{(j)}$ is a {\em primitive} vector
in $\Gs_j$, for each $\,j=1,\ldots,r$. This means that we limit our
analysis of CQHLs to what we call {\em primitive} CQHLs; see
the definition after~\cfr{Qd}.


\begin{flushleft}
\section{Basic Invariants and Elementary Mathematical
Properties of Chiral QH Lattices}
\label{s4}
\end{flushleft}

Let $(\Gamma, \Q)$ be a {\em chiral\,} QH lattice (CQHL), i.e., a QH
lattice describing (a universality class of) QH fluids with edge
currents of a {\em definite chirality}. In this section, we
describe elementary mathematical properties of $(\Gamma, \Q)$. This
is conveniently done in terms of {\em invariants\mini} of $(\Gamma,
\Q)$. Numerical invariants of a CQHL, $(\Gamma, \Q)$, are numbers
which only depend on its {\em intrinsic\mini} properties. They are
{\em independent\,} of the choice of a basis in $\Gamma$ and of a
``reshuffling'' of electric charge assignments corresponding to a
transformation of $\Q$ by an orthogonal symmetry of $\,\G$.

The two most elementary invariants of an integral lattice $\G$
are its {\em dimension} $N$ and its {\em discriminant}
$\,\D=|\Gs/\,\G|\mini$; see Sect.\,2.

What do we know about the values of $N$ and $\Delta$ occuring
for a CQHL that describes a {\em real\,} QH fluid with impurities?

The honest answer is: not much! Let ${\cal E}_e$ denote the average
energy per electron in the ground state of a QH fluid. Let $\,n_e
\equiv n\,$ denote the electron density, $n_I\,$ the density of
impurities, and ${\cal E}_I$ the average potential energy
corresponding to a single impurity. Then we can form the
dimensionless quantity

\be
\alpha \:=\: {\cal E}_e/({\cal E}_I \cdot \frac{n_I}{n_e})\ ,
\label{alpha}
\ee

\noi
and, at some fixed value of the filling factor $\nu$, the maximal
value, $N_*$, of $N$ is an increasing function of $\alpha\mini$;
$(N_*\propto \alpha)$.

One may argue that the density and strength of impurities and the
Wigner-lattice instability yield upper bounds on the discriminant
$\Delta$ of a CQHL $(\Gamma, \Q)$.

What, as physicists, we are longing for are invariants, $\cal{J}\!$,
of CQHLs with the property that if $(\Gamma, \Q)$ is a CQHL
corresponding to a {\em real\,} QH fluid the values of its
invariants $\,{\cal J}\!={\cal J}\!(\G,\Q)\,$ can either be
constrained by {\em experimental data} or by {\em safe
theoretical arguments}. Such invariants are, for example, the {\em
``relative-angular-momentum invariants''}, $\lm$ and $\,\lM\mini$,
described in~\cite{FT}. We first give a {\em mathematical\,}
definition of these invariants and then explain what {\em
physical\,} quantities they correspond to.

Let $(\Gamma, \Q)$ be a primitive CQHL. Since $\Q$ is a primitive
vector of $\,\Gs$, there is a basis, $\{ \avecd{\q} \}$, of $\,\G$
such that

\ben
q_{el}(\q_i) \:=\:\: \sca{\Q\,}{\q_{i}} \:=\: 1\ , \hfa
 i=1,\ldots,N \ .
\een

\noi
The set of {\em all\,} such bases of $\,\G$ is denoted by $\BQ$.
We define

\be
\Lmg \::=\:
\min_{\q\mini\in\mini\G\mimini,\:\las\Q\mini,\mini\q\ras\mini =
\mini 1} \mini \la\q\,,\q\ra\ ,
\label{Lm}
\ee

\vav

\be
\LMg \::=\: \min_{\{\q_1,\ldots,\mini\q_N \}\in \BQs}
\mini \left( \max_{1\leq i\leq N} \la\q_i,\q_i\ra\right)\ .
\label{LM}
\ee

\noi
By~\cfr{parity}, $\Lmg$ and $\LMg$ are {\em odd}, positive integers,
and $\,\Lmg \leq \LMg\mini$.

Suppose that $(\Gamma, \Q)$ is decomposable (see~\cfr{Gd}) and let

\ben
(\Gamma,\Q) \:=\: \bigoplus_{j=1}^r \, (\Gamma_j, \Q^{(j)})
\een

\noi
be the decomposition of $(\Gamma, \Q)$ into indecomposable CQHLs,
$(\Gamma_j, \Q^{(j)}),\ j=1,\ldots,r$. Note that, since $(\Gamma,\Q)$
has been assumed to be {\em primitive\mini} (see Sect.\,2,
following~\cfr{Qd}), each vector $\Q^{(j)}$ is a (non-zero) {\em
primitive} vector of the lattice $\Gamma^\ast_j$ dual to $\Gamma_j$,
and every sublattice $\Gamma_j$ is {\em odd}, so that $(\Gamma_j,
\Q^{(j)})$ is an indecomposable CQHL. We define the
relative-angular-momentum invariants, $\lm$ and $\,\lM$, by

\be
\lmg \::=\: \min_{1\leq j\leq r} \Lm (\Gamma_j, \Q^{(j)}) \:\geq\:
\Lmg\ ,
\label{lm}
\ee

\vav

\be
\lMg \::=\: \max_{1\leq j\leq r} \LM (\Gamma_j,\Q^{(j)}) \:\geq\:
\LMg\ .
\label{lM}
\ee

We pause to explain the physical meaning of the invariants
$\Lm$ and $\LM$. For this purpose, we consider a state of the
system where two quasi-particles, with quantum numbers
corresponding to vectors $\q_1$ and $\q_2$ in $\mini\Gp$ and
localized near two points $\vec{x}_1 \ne \vec{x}_2$ in the plane of
the system, are created {}from the ground state. {}From
Chern-Simons theory and its relation to chiral conformal field
theory it is known that the physical state vector,
$\Psi=\Psi(\vec{x}_1, \q_1;\vec{x}_2, \q_2)$, is then given by

\be
\Psi(\vec{x}_1, \q_1;\vec{x}_2, \q_2) \:=\:
(z_1-z_2)^{\las\q_1,\mini\q_2\ras} \,\Phi(\vec{x}_1, \q_1;\vec{x}_2,
\q_2)\ ,
\label{CB}
\ee

\noi
where $z=x+iy$ is the complex number corresponding to a point
$\vec{x}=(x, y)$ in the plane of the system, and where
$\Phi(\vec{x}_1, \q_1;\vec{x}_2, \q_2)$ is {\em single-valued\,} in
$\vec{x}_1$ and $\vec{x}_2$. Let $\BBd{L}{z}$ denote the component
of the relative-angular-momentum operator along the axis
perpendicular to the plane of the system. Then~\cfr{CB} implies that

\be
\BBd{L}{z} \Psi \:=\:\: \la \q_1\mini, \q_2 \ra\!\Psi +
(z_1-z_2)^{\las\q_1,\mini\q_2\ras}\BBd{L}{z} \Phi\ .
\label{ceLz}
\ee

\noi
Since $\Phi$ is single-valued in $\vec{x}_1$ and $\vec{x}_2$, it
follows {}from~\cfr{ceLz} that the possible eigenvalues of
$\,\BBd{L}{z}$ corresponding to states describing two quasi-particles
with vectors $\q_1$ and $\q_2$ are given by

\ben
\sca{\q_1\mini}{\q_2}\! +\; m\ , \hwh m \in \Z\ .
\een

\noi
If the two quasi-particles are electrons with quantum numbers $\,\q_1
= \q_2 = \q$, where $\q$ is a point of the lattice $\Gamma$ with
$\,q_{el}(\q)=\:\la\Q\,,\q\ra = 1$, then, in states of low energy,
$m$ is non-negative, and it follows that the relative angular
momentum, $L$, of the state vector
$\,\Psi(\vec{x}_1,\q;\vec{x}_2,\q)$ is at least as large as
$\,\la\q\,,\q\ra$, i.e.,

\be
 L \:\geq\:\: \la\q\,,\q\ra \ .
\ee

\noi
Thus, by~\cfr{Lm}, $\Lmg$ is the smallest possible relative angular
momentum of a state describing two electrons excited {}from the
ground state in a QH fluid corresponding to the CQHL $(\Gamma, \Q)$.
For a QH fluid describing a system of non-interacting electrons
filling the lowest Landau level, the Hall conductivity $\sH$ is
unity, and $\,\Lm=\LM=1\mini$. For the basic Laughlin
fluid~\cite{L}, we have that

\ben
\sH \:=\: \frac{1}{3} \ , \hah \Lm \:=\: \LM \:=\: 3\ .
\een

\noi
We shall prove below that, for an arbitrary {QH fluid} described
by a CQHL $(\G,\Q)$,

\be
\LMg \:\geq\: \Lmg \:\geq\: {\la\Q\,,\Q\ra\!}^{-1} \:=\:
\sH^{-1}(\G,\Q)\ .
\label{CauchyS}
\ee

Theoretical arguments and numerical simulations~\cite{LmaxBound}
suggest that the relative-angular-momentum invariant $\LM$ obeys a
universal upper bound

\be
\LM \:\leq\: \Ls \ ,
\label{LBound}
\ee

\noi
with $\Ls= 7$ or $9$. This can also be understood as follows: If
$\LM$ were larger than $7$ or $9$, say, the density of
electrons in the  ground state of a QH fluid corresponding to such a
value of the relative-angular-momentum invariant would be so {\em
small\,} that the system could lower its energy if the electrons
formed a {\em Wigner lattice}~\cite{WignerLatt}. But a Wigner
lattice is {\em not\,} an incompressible state.

By~\cfr{CauchyS}, the bound~\cfr{LBound} implies a lower bound on the
Hall fraction $\sH$ of real QH fluids:

\be
\sH \:\geq\: \frac{1}{\Ls} \:=\: \frac{1}{9} \; \mbox{ or } \;
\frac{1}{7}\ .
\ee

\noi
This lower bound is in agreement with experimental data; see
Fig.\,1.1\mini.

Using Hadamard's inequality for the determinant of the lattice Gram
matrix (see~\cite{FT,Sieg}), it is easy to prove that the
discriminant $\,\D(\Gamma,\Q)=|\Gs/\G|\,$ of a CQHL $(\Gamma,\Q)$
corresponding to a QH fluid is bounded by

\be \D(\Gamma,\Q) \:\leq\: \LMg^N \ ,
\label{DBound}
\ee

\noi
where $N$ is the dimension of $\,\G$ which can be bounded, in
principle, in terms of the quantity $\alpha$ defined in~\cfr{alpha},
i.e., which remains {\em finite} for systems with a positive density
of impurities of finite strength. Then~\cfr{DBound} shows that the
discriminant $\D$ can only take a finite (but possibly large) set of
values. As discussed in~\cite{FT}, this implies that the problem of
classifying all CQHL corresponding to physically realizable QH
fluids is a {\em finite\mini} problem.

Equipped with the invariants $N$, $\D$, $\lm$, and $\,\lM$ of CQHLs,
we can begin to classify such lattices.

First, we prove the bound~\cfr{CauchyS} on $\sH^{-1}$ which holds
for general CQHLs, not necessarily proper or primitive. The
proof is an easy application of the Cauchy-Schwarz inequality,

\ben
\frac{{\la\n\,,\q\ra\!}^2}{\la\n\,,\n\ra} \:\leq\:\: \la\q\,,\q\ra \
, \een

\noi
for arbitrary vectors $\q$ and $\,\n \ne 0\,$ in a vector space
$V$. Setting  $\,\n=\Q\in\!\Gs\subset V\,$ and choosing $\q$ to lie
in $\mini\G$ we conclude that

\bea
\sca{\q\,}{\q} &\geq &
{\sca{\Q\,}{\q}\!}^2 \; {\sca{\Q\,}{\Q}\!}^{-1} \nonumber \\
&=& {\sca{\Q\,}{\q}\!}^2 \; {\sH}^{-1}(\G,\Q)\ .
\label{CSp}
\eea

\noi
Since, for any non-zero vector $\Q$ in $\mini\Gs$, and any vector
$\q$ in $\mini\G$ with $\,q_{el}(\q)=\:\la\Q\,,\q\ra\ne 0$, we have
that $\,{\la \Q\,,\q\ra\!}^2 \geq 1$, \cfr{CSp} implies that

\be
\la\q\,,\q\ra \:\geq\: \sH^{-1}(\G,\Q) \ ,
\label{CSpp}
\ee

\noi
for an arbitrary vector $\q$ in $\mini\G$, with $\,\la\Q\,,\q\ra
\ne 0$. Recalling the definition~\cfr{Lm} of the invariant $\Lmg$,
we find that the bound~\cfr{CSpp} implies~\cfr{CauchyS}. Moreover,
by~\cfrs{lm}{lM}, we have that

\be
\lMg \:\geq\: \lmg \:\geq\: \sH^{-1}(\G,\Q) \ .
\label{lBound}
\ee

In the following, we focus on the classification of CQHLs $(\G,\Q)$
with

\be
\sH(\G,\Q) \:=\:\: \la\Q\,,\Q\ra \:\leq\: 1 \ ;
\ee

\noi
corresponding to a partially or fully filled lowest Landau level. We
divide the half-open interval $\,(0, 1]\,$ into a sequence of
subintervals, or ``{\em windows\,}'', $\Sp$, where

\be
\Sp \::=\: \bigl\{ \sH\ |\ \frac{1}{2\mini p+1} \leq
\sH<\frac{1}{2\mini p-1}\,\bigr\} \ ,
\label{Sp}
\ee

\noi
for $\,p=1,2,\ldots\,$, and $\,\Sigma_0:=\left\{ \sH=1 \right\}$.

In each such window, we attempt to classify all those CQHLs which are
{\em \Lmini\,} in the sense of the following definition.

\proclaim\indent Definition. A CQHL $\,(\G,\Q)$ with $\,\sH(\G,\Q)=
\:\la\Q\, ,\Q\ra\:\in\!\Sp$, $\,p=1,2,\ldots\,$, is {\em \Lmini\,} if
and only if it is {\em primitive} and
\be
\lMg \:=\: 2\mini p+1 \ .
\label{L-min}
\ee

Note that $2\mini p+1$ is the {\em smallest\,} possible value of the
invariant $\,\lM$ that is compatible with the bound~\cfr{lBound} when
$\sH$ belongs to the window $\Sp\mini$. It then follows
{}from~\cfr{lBound} and~\cfr{L-min} that, for an \Lmini\ CQHL
$(\G,\Q)$ with $\,\sH \in\!\Sp$

\be
\lMg \:=\: \lmg \:=\: \LMg \:=\: \Lmg \:=\: 2\mini p+1 \ ,
\label{llLLp}
\ee

\noi
i.e., {\em all\,} relative-angular-momentum invariants take the
smallest possible value compatible with the value of $\sH$.

\proclaim\indent Proposition. Any proper CQHL $(\G,\Q)$ with
$\,\sH(\G,\Q) \in\!\Sp\,$ and $\,\LMg=2\mini p+1,\ p=1,2,\ldots\,$,
is a primitive CQHL with $\,\lMg =\LMg = 2\mini p+1$, i.e., it is
{\em \Lmini}. Moreover, if $\,\sH(\G,\Q) < 2/3$, it is {\em
indecomposable}.

\noi
{\em Proof.} The proposition is obviously true if the lattice $\G$ is
indecomposable, since then it is necessarily primitive and, by
definition~\cfr{lM}, $\lMg =\LMg$.

If the lattice is decomposable, i.e.,

\bea
\G & = & \G_1 \oplus \cdots \oplus\,\G_r\ ,\hsp\mbox{and}\nonumber\\
\Gs & = & \Gs_1 \oplus \cdots
\oplus\,\Gs_r\ ,
\label{pdeco}
\eea

\noi
for some $\,r>1$, the vector $\,\Q\in\!\Gs$ has a non-vanishing
projection, $\Q\!\mid_{\Gs_i}$, along each orthogonal summand $\Gs_i$
of the dual lattice $\Gs$, $\,i=1,\ldots,r$, since $(\G,\Q)$ is
proper by hypothesis (see~\cfr{Qd}). The projections
$\Q\!\mid_{\Gs_i}$, however, are not necessarily primitive vectors,
i.e., there are strictly positive integers $\,q_i = \gcd
(\Q\!\mid_{\Gs_i})\,$ and primitive vectors $\Q^{(i)}$ in
$\Gs_i$ such that

\be
\Q \:=\: q_1 \, \Q^{(1)}  +  \cdots  + q_r \, \Q^{(r)}\ ,
\label{Qdp}
\ee

\noi
with $\,\gcd (q_1, \ldots , q_r) = 1$, since $\Q$ is primitive.

By hypothesis, we have that $\,\sH \in\!\Sp\,$ and $\,\LM=2\mini
p+1\mini$; hence~\cfr{CauchyS} gives

\ben
\LMg \:=\: \Lmg \:=\: 2\mini p+1 \ ,
\een

\noi
which implies that there is a {\em basis} for $\mini\G$ made of
vectors $\e$ satisfying

\be
\sca{\e\,}{\e} \:=\: 2\mini p+1\ , \hah \sca{\Q\,}{\e} \:=\: 1 \ .
\label{eqe}
\ee

\noi
Let

\be
\e  \:=\: \e^{(1)}  +  \cdots  +  \e^{(r)}
\ee

\noi
be their orthogonal decomposition according to~\cfr{pdeco}. Then,
by~\cfr{eqe},

\be
1 \:=\: \sum_{i=1}^r \, q_i \sca{\Q^{(i)}}{\e^{(i)}}\ ,
\ee

\noi
and, for the Hall fraction $\sH$, the decomposition~\cfr{Qdp}
implies

\bea
\sH(\G,\Q) &=& \sca{\Q\,}{\Q} \:=\: \sum_{i=1}^r\,q_i^2
\,\sigma_{(i)}\ , \hsp\mbox{with} \nonumber \\
\sigma_{(i)} &:=& \sca{\Q^{(i)}}{\Q^{(i)}} \;(>\:0)\ ,\hsp i =
1,\ldots,r\ .
\label{psum}
\eea

In order to prove the first part of the proposition($(\G,\Q)$
is primitive) we have to show that all $q_i\mini$ in~\cfr{Qdp} equal
unity.

Since the electric charge of any one of the above basis vectors
$\e$ equals unity, $\,q_{el}(\e)=\:\mbox{$\sca{\Q\,}{\e}$}=1$, at
least one of its projections, say $\,\e^{(I)} \in\!\G_I$, must carry
an {\em odd\,} charge, i.e., $\sca{\Q\,}{\e^{(I)}} = q_I
\sca{\Q^{(I)}}{\e^{(I)}}\,$ is an {\em odd\,} integer. Hence
$\,q_I\,$ and $\,\sca{\Q^{(I)}}{\e^{(I)}}\,$ are both {\em odd}, and
$\,\sca{\e^{(I)}}{\e^{(I)}}\,$ must be {\em odd}, too,
by~\cfr{parity}). Thus, we have found a vector $\e^{(I)}$ in the
euclidean lattice $\,\G_I\subset\G\,$ whose squared length satisfies

\be
\sca{\e^{(I)}}{\e^{(I)}} \:\leq\:\: \sca{\e\,}{\e} \:=\: 2\mini p+1
\ ,
\label{eI}
\ee

\noi
(equality holding only if $\,\e^{(I)} = \e$), and whose charge
$\,q_{el}(\e^{(I)})=\:\sca{\Q^{(I)}}{\e^{(I)}}$ is non-vanishing
and odd.

Next, given the decomposition~\cfr{psum} and the assumption that
$\,\sH \in\!\Sp$, the value of $\,\sigma_{(I)} =\:
\sca{\Q^{(I)}}{\Q^{(I)}}$ is bounded, and we have

\be
\sigma_{(I)}^{-1} \:>\: q_I^2 \; (2\mini p-1)\ ,
\label{sii}
\ee

\noi
the inequality being {\em strict\,} because $\,r>1\,$ and {\em no}
summand $\,q_i^2\,\sigma_{(i)}\,$ is vanishing in~\cfr{psum}.
Inequality~\cfr{sii} and the Cauchy-Schwarz inequality~\cfr{CSp} for
the pair $(\G_I,\Q^{(I)})$ then imply

\ben
\sca{\e^{(I)}}{\e^{(I)}} \:\geq\: {\sca{\Q^{(I)}}{\e^{(I)}}\!}^2 \;
\sigma_{(I)}^{-1} \:>\: q_I^2 \; (2\mini p-1) \ .
\een

\noi
This inequality, however, contradicts inequality~\cfr{eI} unless

\ben
q_I \:=\: 1\ , \hah \e^{(I)}  = \e\ .
\een

\noi
Hence the basis vector $\e$ necessarily lies entirely in one
orthogonal summand of the decomposition~\cfr{pdeco}; and since there
is a {\em basis\mini} of $\,\G$ of such vectors $\e$, we conclude
that in~\cfr{Qdp}

\ben
q_i \:=\: 1\ , \hfa i=1,\ldots,r\ .
\een

\noi
This proves that $(\G,\Q)$ is primitive. Furthermore, for the
relative-angular-momentum invariant $\,\lMg$ (see~\cfr{lM}), the
above reasoning implies that

\ben
\lMg \:=\: \LMg \:=\: 2\mini p+1\ ,
\een

\noi
i.e., the CQHL $(\G,\Q)$ is \Lmini.

We can now easily prove the second part of the proposition about the
indecomposablility of \Lmini\ CQHLs with $\,\sH<2/3\mini$. Let
$(\G,\Q)$ be such a lattice with $\,\sH\in\!\Sp$, $\,p=1,2,\ldots\ $.
Assuming $\,r>1$ in~\cfr{pdeco}, we have proven that, for each
summand $\Gamma^{(j)}$ of $\,\G$, $\,j=1,\ldots,r$, there is a basis,
$\{\e_1^{(j)},\ldots, \e_{s_j}^{(j)}\}$, of minimal-length vectors
with charge $1$, i.e.,

\be
\la\e_{k_j}^{(j)},\e_{k_j}^{(j)}\ra \:=\: 2\mini p + 1\ , \hah
\la\Q^{(j)},\e_{k_j}^{(j)}\ra \:=\: 1\ ,\hsp k_j=1,\ldots,s_j\ .
\label{ek}
\ee

\noi
Moreover, for the Hall fraction $\sH$, we have that

\ben
\frac{1}{2\mini p+1}Ê\:\leq\: \sH(\G,\Q) \:=\: \sigma_{(1)} + \cdots
+ \sigma_{(r)} \:<\: \min\,\{\,\frac{1}{2\mini p-1}\,,\,\frac{2}{3}
\:\}\ .
\een

\noi
Then, by the assumption that $\,r>1$, at least one component, say,
$(\G_I,\Q^{(I)})$ satisfies

\be
\sigma_{(I)} \:=\:\: \la\Q^{(I)},\Q^{(I)}\ra \:<\:
\min\, \{\, \frac{1}{2}\,\frac{1}{2\mini p-1}\,,\,\frac{1}{3}\,\}\ ,
\label{sIp}
\ee

\noi
since $\sigma_{(j)} \neq 0$, for all $\,j=1,\ldots,r$. Applying again
the Cauchy-Schwarz inequality~\cfr{CSp} to the pair
$(\G_I,\Q^{(I)})$, we find, with~\cfr{sIp}, that

\be
\la\e_{k_I}^{(I)},\e_{k_I}^{(I)}\ra \:\geq\:  \sigma_{(I)}^{-1}
\:>\: \max\,\{\,4\mini p-2\,,\, 3\,\}\ ,
\ee

\noi
which contradicts the equality in~\cfr{ek}. Hence, for an \Lmini\
CQHL $(\G,\Q)$, $\,\sH(\G,\Q)<2/3\,$ implies $\,r=1$, which means
that the lattice $\G$ is indecomposable. \rule{2.5mm}{2.5mm}

\vspace{5mm}

Note that the above bound for the value of $\sH$ of indecomposable
\Lmini\ CQHLs is optimal: It is very easy to construct a {\em
composite} \Lmini\ CQHL at the threshold value $\,\sH = 2/3\mini$.
E.g., one takes the direct sum, $\,\Gamma = \Gamma_1 \oplus
\Gamma_2$, and $\,\Q=\Q^{(1)}+\Q^{(2)}$, of two one-dimensional
CQHLs with $\,\sigma_{(1)} = \sigma_{(2)} = 1/3$, i.e.,
$\G_i=\sqrt{3}\,\Z\,$ is generated by $\e^{(i)}$ with squared length
$\,\la\e^{(i)},\e^{(i)}\ra=3$, and $\Q^{(i)} = \es^{(i)} =
\frac{1}{3} \, \e^{(i)}$ is the dual basis vector in $\Gs_i$,
$i=1,2\mini$; $\,\sH = \sigma_{(1)} + \sigma_{(2)} = 2/3$, and the
lattice $\G$ is primitive and \Lmini. The two one-dimensional CQHLs
$(\G_i,\Q^{(i)}),\ i=1,2$, correspond to the basic Laughlin
fluid~\cite{L} at $\,\sH=1/3\mini$.

In order to state some powerful general classification results on
\Lmini\ CQHLs with $\sH < 1$ (Thms.\,4.1 and\,4.2 below), we
introduce a family of maps $\,\Sm,\ p=1,2,\ldots\,$, between CQHLs of
equal dimension. As we will see shortly, acting with $\Sm$ on a
CQHL $(\G,\Q)$ simply shifts the inverse Hall fraction
$\sH^{-1}(\G,\Q)$ and the relative-angular-momentum invariants
$\LMg$ and $\Lmg$ by a common even integer $2\mini p$.

\proclaim\indent Definition. For any positive integer $p$, the {\em
shift map} $\Sm$ is a map between proper CQHLs of equal dimensions,
$\Sm \mini:\, (\G,\Q) \mapsto (\Gamma^\prime,\Q^\prime)$. {}From an
arbitrary basis $\{ \f_1,\ldots,\f_N \}$ of $\,(\G,\Q)$, a basis
$\{\f_1^\prime ,\ldots, \f_N^\prime \}$ and a charge vector
$\Q^\prime$  of the image $(\G^\prime,\Q^\prime)$ is constructed such
that
\be
\la\f_i^\prime,\f_j^\prime\ra \:=\:\: \la\f_i,\f_j\ra +\;2\mini p
\la\f_i,\Q\ra \la\Q\,,\f_j\ra\ ,
\label{sm1}
\ee
and
\be
\la\Q^\prime,\f_j^\prime\ra \:=\:\: \la\Q\,,\f_j\ra\ .
\label{sm2}
\ee

To see that this definition makes sense, i.e., that $(\G^\prime,
\Q^\prime)$ is a well-defined proper CQHL, take any so-called
``normal'' basis $\{\q,\e_2,\ldots,\e_N\}$ of $\,\G$ which is
characterized by $\,\la\Q\,,\q\ra = 1\,$ and $\,\la\Q\,,\e_i\ra=0$,
for $i= 2, \ldots,N$. In such a basis, the map $\Sm$ ``shifts'' the
squared length of the vector $\q$ by the even integer $2\mini p$,

\ben
\la\q^\prime,\q^\prime\ra \:=\:\: \la\q\,,\q\ra +\; 2\mini p\ ,
\een

\noi
and the neutral sublattice,

\be
\G_0(\G,\Q) \::=\: \{\, {\bf v} \in
\Gamma  \; | \; \la\Q\,,{\bf v}\ra = 0 \,\}\ ,
\ee

\noi
(generated by the vectors $\e_i,\ i=2,\ldots,N$) is left unchanged,
i.e.,

\be
\G_0(\G^\prime,\Q^\prime) \:=\: \Gamma_0(\G,\Q)\ .
\label{neut}
\ee

Given~\cfr{neut}, and defining $\,\gamma(\G,\Q):=
\det(K\!\mid_{\Gamma_0}) =\widetilde{K}_{11}\ (>0)$, where
$\widetilde{K}$ is the cofactor matrix of the Gram matrix $K$ of the
lattice $\mini\Gamma$ which is associated to the normal basis above,
one finds that

\be
\gamma(\G^\prime,\Q^\prime) \:=\: \gamma(\G,\Q)\ .
\label{gprime}
\ee

\noi
Furthermore, computing the lattice discriminant
$\,\D(\G^\prime,\Q^\prime) := \det(K^\prime)$ of $\,\G^\prime$
generated by $\{\q^\prime, \e_2,\ldots,\e_N\}$, we find that

\be
\Delta(\G^\prime,\Q^\prime) \:=\: \Delta(\G,\Q) + 2\mini p
\,\gamma(\G,\Q)\ .
\label{Dprime}
\ee

\noi
Eqs.~(\ref{gprime}) and~(\ref{Dprime}) show that $\G^\prime$ is a
euclidean lattice which, by~\cfr{sm1}, is odd. Primitivity of
$\Q^\prime$ results form that of $\Q$ through~\cfr{sm2}. Moreover,
since any improper orthogonal component of a CQHL necessarily lies in
its neutral sublattice (see~\cfr{Qd}), the image $(\G^\prime,
\Q^\prime)$ is a proper CQHL if $(\G,\Q)$ is proper.

Since the Hall fraction $\sH$ can be written as

\be
\sH(\G,\Q) \:=\:\: \la\Q\,,\Q\ra \:=\: ( K^{-1})_{11} \:=\:
\frac{\widetilde K_{11}}{\D(\G,\Q)} \:=\:
\frac{\gamma(\G,\Q)}{\D(\G,\Q)}\ ,
\label{sHgD}
\ee

\noi
the change in $\sH$, when acting with the shift map $\Sm$, is given
by

\be
\sH^{-1}(\G^\prime,\Q^\prime) \:=\: \sH^{-1}(\G,\Q) + 2\mini p\ .
\label{smsH}
\ee

\noi
(Here the reader may recognize the ``$D$-operation'' (or ``first
move'') of the Jain-Goldman hierarchy scheme~\cite{JG,Jain}). Note
that~\cfr{smsH} implies that all CQHLs $(\G^\prime,\Q^\prime)$
obtained through the action of shift maps $\Sm,\ p=1,2,\ldots\,$,
necessarily have

\be
\sH(\G^\prime,\Q^\prime) \:<\: \frac{1}{2\mini p}\ .
\ee

In~\cfr{sHgD}, the quantities $\gamma$ and $\Delta$ are obviously
not necessarily coprime integers. Thus, defining by $\,l(\G,\Q) :=
\gcd(\Delta, \gamma)\,$ the ``level'' of a CQHL $(\G,\Q)$
(see~\cite{FT}), and denoting by $\nH$ and $\dH$ the numerator and
denominator, respectively, of the Hall fraction, i.e.,
$\,\sH=\nH/\dH$, one has $\,\nH=\gamma/l$, and $\,\dH=\Delta/l$.
Applying the shift map $\Sm$, we find {}from~\cfrs{gprime}{Dprime}
that $\,l(\G^\prime,\Q^\prime)=\gcd(\Delta^\prime,\gamma^\prime)=
\gcd(\Delta + 2\mini p\,\gamma, \gamma) = \gcd(\Delta, \gamma) =
l(\G,\Q)$, and hence

\be
\nH(\G^\prime,\Q^\prime) \:=\: \nH(\G,\Q) \ ,\hah
\dH(\G^\prime,\Q^\prime) \:=\: \dH(\G,\Q) + 2\mini p\,\nH(\G,\Q)\ .
\ee

Starting again {}from a normal basis, the change of the
relative-angular-momentum invariants $\Lm$ and $\LM$, defined
in~\cfrs{Lm}{LM}, is given by simply shifting them by $2\mini p$,

\be
\Lm(\Gamma^\prime,\Q^\prime) \:=\: \Lm(\Gamma, \Q)+2\mini p\ , \hah
\LM(\Gamma^\prime,\Q^\prime) \:=\: \LM(\Gamma,\Q)+2\mini p\ .
\label{smL}
\ee

We emphasize that this simple transformation rule does {\em not\,}
hold, in general, for the invariants $\,\lm$ and $\,\lM$ defined
in~\cfrs{lm}{lM}. This fact points to a basic pitfall one has to
avoid when using the shift maps $\Sm$ in the classification of CQHLs:
the maps $\Sm$ do not necessarily preserve the decomposability
properties of CQHLs. Moreover, they do not, in general, preserve the
primitivity property that we have required for physically relevant
{\em composite} CQHLs (see the end of Sect.\,2); a CQHL that is not
primitive can have as its image under $\Sm$ a primitive CQHL.

We illustrate this situation by a pair of CQHLs in two dimensions for
which we give the corresonding Gram matrices, $\,K^{(\prime)}_{ij}
=\:\sca{\q_i^{(\prime)}}{\q_j^{(\prime)}}\!$, in symmetric bases, $\{
\q_1^{(\prime)},\q_2^{(\prime)} \}$, where
$\,\sca{\Q^{(\prime)}}{\q_i^{(\prime)}}=1$, for $\,i=1,2\mini$:

\be
K \:=\: \left( \begin{array}{cc}
1  & -1 \\
-1 & 3
\end{array} \right) \; \mbox{ at } \; \sH \:=\: 3 \quad
\stackrel{\textstyle{\cal S}_1}{\longmapsto}\quad
K^\prime \:=\: \left( \begin{array}{cc}
3 & 1 \\
1 & 5
\end{array} \right) \; \mbox{ at } \; \sH \:=\: \frac{3}{7}\ .
\label{Sexamp}
\ee

\noi
One then easily checks that the preimage lattice described by $K$ is
decomposable according to $\,\Gamma=\G_1\oplus\G_2 \simeq \Z
\oplus \sqrt{2}\,\Z\,$, where the first summand is generated by
$\e^{(1)}$ with $\,\sca{\e^{(1)}}{\e^{(1)}}=1$, and the second one
by $\e^{(2)}$ with $\,\sca{\e^{(2)}}{\e^{(2)}}=2\mini$. Moreover,
the decomposition of the charge vector $\Q$ corresponding to
$\,\Gs=\Gs_1\oplus\Gs_2\,$ reads $\,\Q=\es^{(1)}+2\mini\es^{(2)}$,
where the dual basis in $\Gs$ is given by $\,\es^{(1)}=\e^{(1)}\,$
and $\,\es^{(2)}=\e^{(2)}/2\mini$. Thus, for the restriciton of
$\Q$ to the second summand, $\Q\!\mid_{\Gs_2}=2\mini\es^{(2)}$, we
find that $\,\gcd(\Q\!\mid_{\Gs_2}) =
\gcd(\sca{2\mini\es^{(2)}}{\e^{(2)}})=2$, and the composite CQHL
$(\G,\Q)$ is not primitive. Physically, the second summand in the
decomposition of $(\G,\Q)$ corresponds to a QH subfluid at
$\,\sH=2\,$ that consists of a bosonic charge$\mini$-2 condensate in
which there are no elementary electrons. Finally, the image CQHL
$(\G^\prime,\Q^\prime)$ specified by $K^\prime$ in~\cfr{Sexamp} can
be shown to be indecomposable and hence primitve.

{}From the discussion above it follows that the shift maps $\Sm,\
p=1,2,\ldots\,$, are injective on the set of proper CQHLs and that
they are surjective onto the subsets of proper CQHLs with
$\,\sH<1/(2\mini p)$. Hence they establish interesting {\em
bijections\mini} between subsets of {\em proper CQHLs} at
corresponding ``shifted'' values of the inverse Hall fraction
$\sH^{-1}$ (see~\cfr{smsH}) and of the relative-angular-momentum
invariants $\Lm$ and $\LM$ (see~\cfr{smL}).

Although such bijections do {\em not\,} hold, {\em in general}, on
the physically relevant restricted subset of {\em primitive\mini}
CQHLs, they {\em hold\,} agian on the more restricted subset of
{\em \Lmini\,} CQHLs (see~\cfr{L-min}), and two powerful, general
classification results can be established there.

We define the following classes of \Lmini\ CQHLs with Hall fractions
$\,\sH<1\mini$:

\bea
\Hp &:=& \{\, (\G,\Q)\,\mbox{ a CQHL}\ |\ \sH(\G,\Q)\in\!\Sp\,\mbox{
and \Lmini, i.e., $(\G,\Q)$ is primitive} \nonumber \\
 & & \mbox{ and }\,\lmg=\lMg =2\mini p+1\,\}\ ,\hsp\ p=1,2,\ldots\ ,
\label{Hp}
\eea

\noi
where the windows $\Sp$ of Hall fractions in the interval
$\,(0,1]\,$ have been defined in~\cfr{Sp}.

Recalling (i)~the proposition stating that the set of \Lmini\
CQHLs coincides with the set of all proper CQHLs with minimal value
of $\LM$ consistent with the value of the Hall fraction $\sH$
(see~\cfr{CauchyS}), and (ii)~the ``shifiting'' properties of
the invariants $\LM$ and $\sH$ under the shift maps $\Sm,\
p=1,2,\ldots\,$ (see~\cfrs{smL}{smsH}), we are led to the
following structural result.

\proclaim\indent Bijection Theorem. The sets $\,{\cal H}_p,\
p=2,3,\ldots\,$, of \Lmini\ CQHLs with $\,\sH\in\!\Sp\,$ are
in {\em one-to-one correspondence} with the set ${\cal H}_1$. The
corresponding bijections are realized by the shift maps $\,{\cal
S}_{p-1}: {\cal H}_1 \rightarrow\,{\cal H}_p\mini$.

Our second result goes much beyond this bijection theorem: namely,
on half of each window $\Sp$, the class ${\cal H}_p,\
p=1,2,\ldots\,$, can be determined completely. Let each window $\Sp$
can be split into two subwindows by its mid value $1/(2\mini p)$,
i.e., we define

\be
\Sp^+ \::=\: \{\, \sH \,|\ \frac{1}{2\mini p+1} \leq \sH <
\frac{1}{2\mini p} \}\ ,
\ee

\vav

\be
\Sp^- \::=\: \{\, \sH \,|\ \frac{1}{2\mini p} \leq \sH <
\frac{1}{2\mini p-1} \}\ ,\hsp p=1,2,\ldots\ .
\ee

\proclaim\indent Uniqueness Theorem. The sets $\,{\cal H}_p^+\subset
{\cal H}_p\,$ of all \Lmini\ CQHLs with $\sH \in\!\Sp^+,\
p=1,2,\ldots\,$, coincide with the infinite series $(N=1,2,\ldots\,)$
of indecomposable, $N$-dimensional, maximally symmetric CQHLs with
$SU(N)$-symmetry of $N$-ality $1$, meaning that the elementary
charge$\mini$-1 fermions (electrons) described by these CQHLs
transform under the fundamental representation of $SU(N)$. The
corresponding Hall fractions are
\be \sH \:=\: \frac{N}{2\mini
pN+1},\hsp p,N=1,2,\ldots\ .
\label{UTsHs}
\ee
(In the notation of~\cite{FT,FSTL,FST},
\be
{\cal H}_p^+ \:=\: \{ \: (\mini 2\mini p+1\,|\,^1\mA A_{N-1})\ |\
N=1,2,\ldots\: \}\ , \hfh p=1,2,\ldots\ .)
\ee

Before proving this theorem, we make a few remarks. First, note
that each Hall fraction $\,\sH=N/(2\mini pN+1),\ p,N=1,2,\ldots\,$,
that appears in~\cfr{UTsHs} is realized by a {\em unique\,} element
of these $su(N)$-series of \Lmini\ CQHLs.

Second, since the level $l\mini$ of all the CQHLs given in the
uniqueness theorem equals unity, there holds a {\em charge-statistics
relation\mini} for the quasi-particle excitations of the
corresponding QH fluids; see~\cite{FT} and~\cite{FST}.

Third, the above $su(N)$-series of \Lmini\ CQHLs have already been
discussed in~\cite{FZ} and, following the approach of~\cite{Read},
they can be shown to describe the same states that have been
proposed by the Haldane-Halperin hierarchy scheme~\cite{HH} as well
as by Jain's scheme~\cite{Jain,JG} at the corresponding  Hall
fractions $\sH$. More details on these equivalences are given
in~\cite{FST}, in particular, in Appendix~E. Furthermore, it has been
shown in~\cite{FZ} (see also~\cite{FSR,FT}) that the excitations
(quasi-particles) of the low-energy spectrum of the corresponding QH
fluids carry a representation of the Kac-Moody algebra \suhN\ at
level $1$.

We now turn to the proof of the uniqueness theorem.
\rule[-4mm]{0mm}{5mm}

\noi
{\em Proof.} Let $(\G^\prime,\Q^\prime)$ be any CQHL in ${\cal
H}_p^+$. Since it is \Lmini, we have that $\,\lM(\G^\prime,
\Q^\prime)=\LM(\G^\prime, \Q^\prime)=2\mini p+1\mini$. Moreover,
since $(\G^\prime,\Q^\prime)$ is proper and has $\,\sH<1/(2\mini p)$,
we can act with the inverse shift map $\,\Sm^{-1}={\cal S}_{-p}\,$
on it and get as preimage a proper CQHL, $(\G,\Q)$, of equal
dimension with $\,\LM(\G,\Q)=1\mini$. This implies that there
is a lattice basis, $\{ \e_1,\ldots,\e_N \}$, of $\,\G$ consisting
of charge$\mini$-1 vectors, $\,\la\Q\,,\e_i\ra= 1$, for all
$\,i=1,\ldots,N$, with unit squared length, $\,\la\e_i,\e_i\ra=1$,
for all $\,i=1, ..., N$. Thus, the lattice $\G$ of the preimage has
to be the $N$-dimensional unit hypercubic lattice,

\be
\G \:\simeq\: {\Z}^N \:\simeq\: \Z \oplus \cdots \oplus \Z\ ,
\label{UTG}
\ee

\noi
where each $\e_i$ generates a one-dimensional unit lattice
$\G_i\simeq \Z\mini,\ i=1,\ldots,N$.

The decomposition of the charge vector $\Q$ corresponding
to~\cfr{UTG} can be written in the general form

\ben
\Q \:=\: q_1 \, \Q^{(1)}+ \cdots +q_N \, \Q^{(N)}\ ,
\een

\noi
where none of the $\,q_i:=\gcd(\Q\!\mid_{\G_i}),\ i=1,\ldots,N$,
vanishes by the condition that $\,\G^\prime$, and hence $\G$, be
proper, and each $\Q^{(i)}$ is a primitive vector in $\,\G_i\simeq
{\Z}^\ast \simeq \Z\,$ (i.e., $\Q^{(i)}= \pm \e^{(i)})$,
$i=1,\ldots,N$; see also~\cfr{Qdp}. Next, we show that $\Q$ is
actually completely fixed by the mere existence of a symmetric basis
$\{ \e_1,\ldots,\e_N \}$ for $\,\G$. The fact that

\ben
1 \:=\: q_{el}(\e_i) \:=\:\: \la\Q\,,\e_i\ra \:=\: q_i\la
\Q^{(i)},\e_i\ra\ , \hfa i=1,\ldots,N \ ,
\een

\noi
implies that $\,q_i=1\,$ and $\,\Q^{(i)}=\e_i$, for each
$\,i=1,\ldots,N$. Hence

\be
\Q \:=\: \Q^{(i)}+\cdots + \Q^{(N)} \:=\: \e_1+ \cdots + \e_N\ ,
\ee

\vav

\be
\sH(\G,\Q) \:=\:\: \la\Q\,,\Q\ra \:=\: 1+\cdots +1 \:=\: N \ .
\label{UTsH}
\ee

Thus, by Eqs.~\cfr{UTG}--\cfr{UTsH}, we conclude that the preimage
$(\G,\Q)$ of the \Lmini\ CQHL $(\G^\prime,\Q^\prime)$ is completely
fixed by its dimension $\,N=1,2,\ldots\ $. Applying the shift map
$\Sm,\ p=1,2,\ldots\,$, to $(\G,\Q)$, we obtain the desired result
of a unique $N$-dimensional CQHL $(\G^\prime, \Q^\prime)$ in ${\cal
H}_p^+$ with Hall fraction $\,\sH=N/(2\mini pN+1),\ N,p=1,2,\ldots\
$. More explicitly, in a symmetric basis, $\{
\f_1^\prime,\ldots,\f_N^\prime\}$, with $\,\la\Q^\prime,
\f_i^\prime\ra=1$, for $\,i=1,\ldots,N$, the Gram matrix $K^\prime$
of $\,\G^\prime$ reads

\be
K^\prime_{ij} \::=\:\: \la\f_i^\prime,\f_j^\prime\ra \:=\:
\delta_{ij} + 2\mini p\ .
\label{suNp}
\ee

\noi
Following the transformation steps given in~\cite{FZ}, we can finally
make the presence of the (global) $SU(N)$-symmetry exhibited by
$(\G^\prime,\Q^\prime)$ and encoded in~\cfr{suNp} more transparent.
Choosing to a suitable normal basis $\{\q^\prime,\e^\prime_2,\ldots,
\e^\prime_N\}$, where $\,\la\Q^\prime,\q^\prime\ra=1$, and
$\,\la\Q^\prime,\e_i^\prime\ra =0$, for $\,i=2, \ldots,N$, the
associated Gram matrix $K_n^{\prime}$ (which is eqivalent to
$K^\prime$ in~\cfr{suNp}) reads

\be
K_n^{\prime} \;=\;
\left( \begin{array}{cccccc}
2p+1 & -1 &  0 &  0 & \cdots &  0 \\
 -1  &  2 & -1 &  0 & \cdots &  0 \\
  0  & -1 &  2 & -1 & \cdots &  0 \\
  0  &  0 & -1 &  2 & \cdots &  0 \\
\vdots&\vdots&\vdots&\vdots&\ddots&\vdots \\
  0  &  0 &  0 &  0 & \cdots &  2
\end{array} \right)\ ,
\ee

\noi
The lower-right, $(N\!-\!1)$-dimensional block is recognized to be
the Cartan matrix of $\,A_{N-1}\simeq su(N)$, the Lie algebra of the
(global) symmetry group $SU(N)$, i.e., the neutral sublattice
$\G_0(\G^\prime,\Q^\prime)$ of the \Lmini\ CQHL
$(\G^\prime,\Q^\prime)$ in ${\cal H}_p^+$ is isomorphic to the root
lattice of $su(N)$. \rule{2.5mm}{2.5mm}


\begin{flushleft}
\section{Phenomenological Implications}
\end{flushleft}

The purpose of this section is to explore the phenomenological
implications of the theoretical results presented in Sect.\,4 and to
confront them with experimental data. We have focused our attention
on {\em chiral\,} QH lattices with Hall fractions $\,\sH\leq 1\mini$.
We recall that such lattices describe QH fluids that have only
electrons (or holes) as fundamental charge carriers and whose edge
currents are of {\em  definite chirality\mini}. We emphasize that,
when restricting considerations to CQHLs, we do {\em not\,} make use
of the idea of ``charge conjugation'', $\,\sH=1-\sH^\prime$, in the
discussion of QH fluids with $\,1/2<\sH<1\mini$. In this interval,
charge conjugation is usually invoked in other approaches to the
fractional QH effect; see~\cite{HH,Jain,JG}. There is no difficulty,
within our framework, to go beyond the chirality assumption and to
consider the general classification problem of mixed-chirality QH
lattices including, e.g., those corresponding to the QH fluids
proposed by the charge-conjugation picture; see~\cite{FST}. The
{\em general\,} problem, however, is exorbitantly involved, and
conclusions lack the simplicity and transparency of the results
obtained when restricting the analysis to the class of CQHLs.
Experimentally, it would be interesting to test the chirality
assumption by direct edge-current measurements, e.g., of the type
reported in~\cite{Ash}; (which, by the way, are compatible with a
{\em purely chiral\,} structure of the QH fluid at $\,\sH=2/3$
that has been studied there).

\vspace{5mm}

\noi
{\bf General Structuring Results}

\noi
$\bullet\ $ {\em The inequality $\,\lM\geq\sH^{-1}$.} The first
important result of our analysis is that, for an {\em arbitrary\,}
CQHL $(\G,\Q)$, the associated invariants $\sH(\G,\Q)$ and
$\,\lM(\G,\Q)$ satisfy the fundamental inequality $\,\lM(\G,\Q) \geq
\sH^{-1}(\G,\Q)$; see~\cfrs{lm}{lBound}. This result implies a first
{\em organizing principle\,} for CQHLs with $\,0<\sH\leq 1\mini$. It
offers a natural splitting of the interval $\mini(0,1]\mini$ into
successive windows, $\Sp$, defined by $\,1/(2p+1)\leq \sH <
1/(2p-1),\ p=1,2,\ldots\ $. The relative-angular-momentum invariant
$\,\lM$ characterizing an arbitrary CQHL with $\,\sH\in\!\Sp\,$ is
thus greater or equal to $\,2\mini p + 1,\ p=1,2,\ldots\ $. Adopting
the heuristic stability principle of Sect.\,1, CQHLs in successive
windows $\Sp$, for {\em increasing\mini} values of $\mini p\mini$,
are expected to describe QH fluids of {\em decreasing\mini}
stability.

Combining the fundamental inequality above with a (universal)
physical upper bound $\Ls$ on the relative-angular-momentum invariant
$\,\lM$, we conclude that, physically, {\em no} chiral QH fluid can
form with a Hall conductivity  $\,\sH < 1/\Ls$. As mentioned in
Sect.\,4 (see~\cfr{LBound}), theoretical and numerical arguments as
well as the present-day experimental data (see Fig.\,1) suggest that
$\,\Ls=7$. In the sequel, we impose this bound.

Thus, for the classification of {\em physically relevant\,} CQHLs,
we have to consider only the first three windows, $\Sigma_1,\
\Sigma_2$, and $\Sigma_3$, because only there CQHLs with $\,\lM\leq
7\,$ can be found. In particular, in the third window $\Sigma_{3}$,
any physically relevant CQHL $(\G,\Q)$ {\em must\,} have
$\,\lM(\G,\Q) = 7$ meaning that $(\G,\Q)$ has to be {\em
\Lmini\mini}, i.e., all relative-angular-momentum invariants take
their smallest possible value; see~\cfr{llLLp}.
\rule[-4mm]{0mm}{5mm}

\noi
$\bullet\ $ {\em Uniqueness theorem.} The uniqueness theorem of the
previous section classifies all {\em \Lmini\,} CQHLs in the left
halfs, $\Sigp^+$, of the windows $\Sigp$, i.e., in the subintervals
$\,1/(2\mini p+1) \leq \sH<1/(2\mini p),\ p=1,2,\ldots\ $. A unique,
$N$-dimensional, \Lmini\ CQHL is found at every Hall fraction
$\,\sH=N/(2\mini pN+1),\ p,N=1,2,\ldots\ $. Given this uniqueness
theorem and the bound $\,\Ls=7$, we find the following remarkable
result: The classification problem of {\em physically relevant\,}
CQHLs can be {\em completely\,} solved in the small subwindow
$\Sigma_3^+ = [1/7,1/6)$. In $\Sigma_3^+$, the only fractions at
which such lattices can be found are $\,\sH=N/(6N+1),\
N=1,2,\ldots\,$, and the electrons of the corresponding unique,
chiral QH fluids carry an $SU(N)$-symmetry.

In the introduction we have mentioned that any set of CQHLs
satisfying upper bounds on their invariants $\,\lM$ and $N$ is
finite. Here it is interesting to note that, {\em
independently\mini} of any upper bound, $\Ns$, on the dimension $N$,
every {\em closed\,} subinterval in $\Sigma_3^+$ contains only a
{\em finite\mini} number of physically relevant $(\lM\leq 7)$ CQHLs.
Presently we do not know to which extent this is also true in other
(sub)windows. The relevance of this remark derives {}from the fact
that it is rather difficult to find satisfactory bounds $\Ns$ on the
dimension $N$ of physically relevant QH lattices; see the discussion
at the beginning of Sect.\,4\mini. Thus, by-passing the need for a
bound on the dimension would be progress.

So far, the only indication of a QH fluid in $\Sigma_3^+$ has been
found at $\,\sH= 1/7$, corresponding to the first member of this
series. It coincides with the Laughlin fluid at $\,m=7$; see
Sect.\,1\mini. Further probing (although difficult experimentally)
of the subwindow $\Sigma_3^+$ would clearly be interesting!
\rule[-4mm]{0mm}{5mm}

\noi
$\bullet\ $ {\em \Lminiy.} We wish to comment on the {\em assumption
of \Lminiy\,} that we require when trying to classify CQHLs that
correspond to stable physical QH fluids. As discussed above, this
assumption is strictly justified only in the window $\Sigma_{3}$. In
general, it is an implementation of the {\em heuristic stability
principle\mini} described in Sect.\,1 stating that the {\em lower\,}
the values of $\,\lM$ and $N$ of a CQHL, the {\em more stable\,} the
corresponding QH fluid. Thus, the justification of the assumption of
\Lminiy\ in the other two windows, $\Sigma_2$ and $\Sigma_1$, is
directly connected to the validity of this stability principle.
Thanks to the precise predictions it yields, experimental
tests can be proposed to settle its validity. Such tests are
discussed in great detail in~\cite{FST}.

In order to formulate such tests, however, one has to go beyond the
assumption of \Lminiy\ in classifying CQHLs, which requires much
more work and can be achieved, in full generality, only for small
values of the bounds $\Ns$ and $\Ls$. In~\cite{FST}, we have fully
classified all CQHLs with upper bounds on $\,\lM$ and $N$ given by
$\,(\Ls,\Ns)=(7,2)$, $\,(\Ls,\Ns)= (5,3)$, and $\,(\Ls,\Ns)=(3,4)$.
We emphasize that the problem is bound to be very complicated, since
it involves as an input the knowledge of the complete classification
of integral lattices with given bounds on the lattice discriminant,
and this is a very intricate mathematical problem (unsolved, in
general, for euclidean lattices, as needed here). With more patience
and computing skill one might extend the above results slightly (at
most, however, up to cases where $\,\Ns=6$, or $7$) because, in low
dimensions, complete lists of lattices can be
computed~\cite{CSbook,Dick}; see~\cite{FST}. Partial results (for
small discriminants) derived {}from the lattice classification
in~\cite{CSS} have already been given in Table~2 of~\cite{FT}.
Furthermore, we note that there is a natural subclass of CQHLs which
generalize the ones of the uniqueness theorem. They exhibit ``large
symmetries'' and, for that reason, have been called ``maximally
symmetric''. In~\cite{FST}, we have classified all \Lmini,
maximally symmetric CQHLs with $\,\sH\leq 1$, {\em independent\,} of
any upper bound on the dimension $N$.

The upshot of the analysis in~\cite{FST} is that the assumption of
\Lminiy\ for physically relevant CQHLs is well supported by the
presently available experimental data. In the sequel, we adopt it as
our basic working hypothesis.
\rule[-4mm]{0mm}{5mm}

\noi
$\bullet\ $ {\em Bijection theorem.} The second main theorem that
we have proven in the previous section, the bijection theorem,
asserts that there are one-to-one correspondences between the sets
of \Lmini\ CQHLs in the different windows $\Sigp,\ p=1,2,\ldots\ $.
Although this theorem does not classify CQHLs, it is a powerful {\em
structuring device\,} for \Lmini\ CQHLs with $\,\sH\leq 1\mini$. In
particular, it reduces the classification of \Lmini\ CQHLs with Hall
fractions in the entire interval $\,0<\sH<1\,$ to the classification
of such lattices in the ``fundamental domain'' $\,\Sigma_1=[1/3,1)$!

\vspace{5mm}

\noi
{\bf Theoretical Implications versus Experimental Data}

In the remaining part of this section, we shall discuss explicit
consequences of the uniqueness and bijection theorem, complement them
with classification results given in~\cite{FST}, and
confront our conclusions with experimental data, as given in Fig.\,1.
\rule[-4mm]{0mm}{5mm}

\noi
$\bullet\ $ {\em The subwindows $\Sigp^+$.} Adopting the hypothesis
of \Lminiy, the uniqueness theorem tells us that, in the subwindows
$\Sigma_{p}^{+}$, {\em no} (chiral) QH fluids can be found with Hall
conductivities $\,\sH\neq N/(2\mini p N + 1),\ N,p=1,2,\ldots\ $.
Taking a look at Fig.\,1, remarkable agreement between this
theoretical prediction and experimental data is found: QH fluids have
been observed at $\,N/(2N+1),\ N=1, \ldots, 9$, in $\Sigma_{1}^{+}$;
at $\,N/(4N+1),\ N=1,2$, and $3$, in $\Sigma_{2}^{+}$; and, as
already mentioned, at just one value of $\,N/(6N+1)$, namely
$\,N=1$, in $\Sigma_{3}^{+}$. The reader has probably recognized
these Hall fractions as the ones of the ``basic Jain
states''~\cite{Jain}. We repeat that, following the
arguments in~\cite{Read} and~\cite{FZ}, one can show~\cite{FST}
that, at the above fractions, the proposals of the hierarchy
schemes~\cite{HH,JG} and of our \Lmini-CQHL scheme coincide. The
additional insight our approach offers is that all these proposals
have a {\em unique\,} status as \Lmini, chiral QH fluids!

A closer inspection of Fig.\,1 shows that, in the subwindows
$\Sigp^+$, there seems to be {\em only one\mini} fraction,
$\sH=4/11$, at which a weak signal of a QH fluid has been reported,
and which does {\em not\,} belong to the set of fractions described
by the uniqueness theorem. The corresponding experimental data
(reported only once) are somewhat controversial; see~\cite{Hu}.
Theoretically, a QH fluid at $\,\sH=4/11\,$ is predicted by the
Haldane-Halperin~\cite{HH} and the Jain-Goldman~\cite{JG} hierarchy
scheme at {\em low(!)} ``level'' $2$ and $3$, respectively. These
two proposals can be shown~\cite{FST} to belong to the {\em same}
universality class of QH fluids described by a {\em non-\Lmini},
two-dimensional (primitive) CQHL which, in some sense, provides the
``simplest'' example of a non-\Lmini\ CQHL. This fraction marks thus
an interesting plateau value where further experiments might
challenge the hierarchy schemes and/or our working hypothesis of
\Lminiy.

It has been emphasized in the literature (see~\cite{Hu}) that the
{\em absence\mini} in the data of Fig.\,1 of a QH fluid at
$\,\sH=5/13\,$ is quite remarkable. Indeed, this fraction is
conspicuous by its absence {}from the list of observed Hall fractions
(in single-layer systems) with denominator $\,\dH=13$, which are
$\,\sH=3/13,\ 4/13,\ 6/13,\ 7/13,\ 8/13$, and $9/13$. Theoretically,
the Haldane-Halperin~\cite{HH} and the Jain-Goldman~\cite{JG}
hierarchy scheme predict a QH fluid with $\,\sH=5/13\,$ at {\em
low(!)} ``level'' $3$ and $2$, respectively. These two proposals
correspond to a {\em non-chiral\,} QH lattice. We note that, in
addition, there is an (inequivalent) {\em chiral}, but {\em
non-\Lmini\,} QH lattice in three dimensions with $\,\sH=5/13$;
see~\cite{FST}. This fraction is thus another interesting plateau
value where the hierarchy schemes and/or the \Lminiy\ assumption can
be tested further.

By a similar reasoning process, in the first subwindow $\Sigma_1^+$,
all fractions in the open intervals $\,N/(2N+1)<\sH< (N+1)/(2N+3),\
N=1,2,\ldots\,$, are interesting plateau values for testing the
\Lminiy\ assumption. To be explicit, we do {\em not\,} expect stable
QH fluids to form at $\,\sH=4/11(!),\ 5/13(!),\ 6/17,\ 7/19,\
8/21,\ldots\,$ in $\,(1/3,2/5)$, and at $\,\sH=7/17,\ 8/19,\ldots\,$
in $\,(2/5,3/7)$. Note that, with the help of the shift maps ${\cal
S}_1$ and ${\cal S}_2$ (see~\cfr{smsH}), these predictions can
be translated into predictions in the subwindows $\Sigma_2^+$
and $\Sigma_3^+$.
\rule[-4mm]{0mm}{5mm}

\noi
$\bullet\ $ {\em The subwindows $\Sigp^-,\ p=1,2,\ldots\ $.} In the
``complementary'' subwindows, $\Sigp^-$, defined by
$\,1/(2\mini)p\leq\sH<1/(2\mini p-1),\ p=1,2,\ldots\,$, we do
{\em not\,} have a complete classification of \Lmini\ CQHLs.
Nevertheless, we can make interesting observations for these
subwindows by exploiting the bijection theorem of the previous
section and the (partial) classification results given in~\cite{FST}.

First, we note that the experimental data in $\,\Sigma_1^-=
[1/2,1)\,$ (see Fig.\,1) can hardly be interpreted as a complete
``mirror image'' of the data in the interval $\,(0,1/2]$, as one
would expect if charge conjugation were at work {\em in
general\mini}. Second, comparing, the data in the two complementary
subwindows $\Sigma_1^-$ and $\Sigma_1^+$, we find, besides the
prominent series of fractions $\,\sH=n/(2n-1),\ n=2,\ldots,9$,
``mirroring'' the unique fractions in $\Sigma_1^+$ (i.e.,
$\sH=1-\sH^\prime$), data points at $\,\sH=4/5,\ 5/7,\ 7/11,\ 8/11,\
8/13,\ 9/13$, and possibly at $10/17$. This is a first experimental
indication that the sets of QH fluids appearing in the complementary
subwindows $\Sigp^+$ and $\Sigp^-,\ p=1,2,\ldots\,$, are ``{\em
structurally distinct\,}''.

We may ask to which extent the experimental data in Fig.\,1 also
support the one-to-one correspondences predicted by the bijection
theorem between QH fluids in the different subwindows $\Sigp^-$. We
can act ``formally'' with the shift maps ${\cal S}_{p-1},\ p=2\,$ and
$3$, of the bijection theorem on the fractions $\sH$ given in
$\Sigma_1^-$ of Fig.\,1, e.g., ${\cal S}_{p-1}\mini:\mini n/(2n-1)
\mapsto n/(2\mini p\mini n-1)$; see~\cfr{smsH}. The resulting
fractions $\sH$ that we obtain in the two subwindows $\Sigma_2^-$
and $\Sigma_3^-$ are fully {\em consistent\,} with the experimental
data given in Fig.\,1. Experimentally observed are the fractions
$\,\sH=n/(4n-1),\ n=2,3,4$, and $\,\sH=4/13\,$ (very weakly) in
$\,\Sigma_2^- = [1/4,1/3)$, and only one fraction in $\,\Sigma_3^- =
[1/6,1/5)$, namely $\,\sH=n/(6n-1)$, with $\,n=2$.

If the QH fluids appearing in $\Sigma_1^-$ were to correspond to
\Lmini\ CQHLs then, by the logic of the bijection theorem, we would
predict the formation of (chiral) QH fluids at $\,\sH=4/13(!),\
5/17,\ 5/19,\ldots\,$ in $\Sigma_2^-$, and at $\,\sH=3/17,\
4/21,\ldots\,$ in $\Sigma_3^-$. These are thus interesting
plateau-values for experimentation.

What do we know {\em explicitly\mini} about \Lmini\ CQHLs in the
subwindows $\Sigp^-$? As mentioned above, the analysis presented
in~\cite{FST} contains, in particular, a complete classification of
all {\em low-dimensional\,} $(N\leq\Ns=4)$ and of all maximally
symmetric, \Lmini\ CQHLs with $\,\sH\leq 1\mini$.
\rule[-4mm]{0mm}{5mm}

\noi
$\bullet\ $ {\em Summary of results presented in~\cite{FST}
for the fundamental subdomain $\Sigma_1^-$.} The upshot of the
analysis given in~\cite{FST} is that, in $\Sigma_1^-$, natural proposals for
QH fluids at the fractions of the series $\,\sH=n/(2n-1),\
n=2,3,\ldots\,$, are provided by the {\em charge-conjugation
picture}, meaning that the corresponding QH fluids are {\em
composite}. They consist of an electron-rich subfluid with a partial
Hall fraction $\,\sigma_{(1)}=1$, and of a hole-rich subfluid
corresponding to an \Lmini\ CQHL of the $su(N)$-series in
$\Sigma_1^+$ with partial Hall fraction $\,\sigma_{(2)} =-N/(2N+1)$,
where $\,N=n-1\mini$. This is, however, {\em not\,} the full story!
As we have already mentioned in the introduction, it is a {\em
structural property\,} of the subwindows $\Sigp^-$ that, at a given
Hall fraction $\sH$, one typically finds {\em more than one\mini}
\Lmini\ CQHL realizing that fraction. (We recall that this is
much in contrast to the {\em unique\mini} realization of the
fractions $\,\sH= N/(2N+1)\,$ in the complementary subwindow
$\Sigma_1^+$.)

For example, at the {\em finite\mini} series of fractions
$\,\sH=n/(2n-1),\ n=2,\ldots,7$, one finds maximally symmetric,
\Lmini\ CQHLs in dimensions $\,N=10-n\,$ which are based on the root
lattices of the exceptional Lie algebras $E_{\mini 9-n}$, similarly
to the way in which the $su(N)$-QH lattices in $\Sigma_1^+$ are
based on the root lattices of $su(N)$; see the end of Sect.\,4\mini.
While the last two members of this finite series, $\sH=6/11$ and
$7/13$, are realized by unique, low-dimensional ($N=4$ and $3$,
respectively), \Lmini\ CQHLs, the higher dimensional members of this
``$E$-series'' of CQHLs contain several \Lmini, chiral {\em QH
sublattices\mini} of lower dimensions. All these sublattices
(although not necessarily maximally symmetric) represent possible
proposals for QH fluids at the corresponding Hall fractions $\sH$.
They arise naturally {}from symmetry breaking patterns existing
for exceptional Lie groups.

Physically, QH lattice embeddings can describe {\em phase
transitions\,} at a given Hall fraction between ``{\em structurally
different\,}'' QH fluids related by {\em symmetry breaking\mini}.
For example, in~\cite[Appendix D]{FST}, we have found $13$ and $5$
QH sublattices embedded into the QH lattice of the $E$-series at
$\,\sH=2/3\ (E_7)\,$ and $3/5\ (E_6)$, respectively. The {\em
composite\,} \Lmini\ CQHL at $\sH=2/3$ which consists of two
Laughlin subfluids with partial Hall fraction $\,\sigma_{(i)}=1/3,\
i=1,2$, is the lowest dimensional ($N=2$) QH sublattice of the
$E_7$-QH lattice at $\,\sH=2/3$. All the other QH sublattices at
$\,\sH=2/3\,$ are {\em indecomposable}. (Recall that we have proven
in Sect.\,4 that all \Lmini\ CQHLs with $\sH<2/3$ are
indecomposable.) In $\Sigma_1^-$, comples embedding patterns
of \Lmini\ CQHLs are found at the fractions $\,\sH=4/7,\ 5/7,\ 5/9$,
and $1/2\mini$! These fractions are interesting in the light of the
data in Fig.\,1, where phase transitions are indicated at
$\,\sH=2/3,\ 3/5$, and possibly at $5/7$, driven by an added
in-plain component of the external magnetic field
(see~\cite{Cla,Eng,Saj}), and at $\,\sH=2/3$, driven by changing the
density of charge carriers in the system (see~\cite{Eis}); see also
the data reported in~\cite{Su} on phase transitions in
wide-single-quantum-well systems.

(At this point, we remark that by the uniqueness of the \Lmini\ CQHLs
in the subwindows $\Sigp^+$ and our heuristic stability principle,
we do {\em not\,} expect structural phase transitions there. Thus,
what about a possible indication of a magnetic field driven phase
transition at $\,\sH=2/5$? As a matter of fact, in~\cite{FSR} (see
also~\cite{FST}), we have argued that $\,\sH=2/5\,$ is the most
likely plateau value where we may expect a phase transition {}from a
{\em spin-polarized\,} to a {\em spin-singlet\,} QH fluid. While,
structurally, the two phases are described by {\em one and the
same\mini} \Lmini\ CQHL, the phase transition corresponds to a
change {}from an {\em internal\,} $SU(2)$-symmetry to a {\em
spatial\,} $SU(2)_{spin}$-symmetry.)

We complete our short review of results derived in~\cite[see
Fig.\,1.2]{FST} by mentioning that to {\em all\,} data points in
$\,\Sigma_1^-\,$ (including the fraction $\,\sH=1/2$) one can
associate {\em at least one\mini} \Lmini\ CQHL that is either
generic (without special symmetry properties) and low-dimensional
$(N\leq 4)$, maximally symmetric with dimension $\,N\leq 9$ (based
on the root lattice of a simple or semi-simple Lie algebra), or
charge-conjugated to an $su(N)$-lattice in $\Sigma_1^+$. Within
these three subclasses of CQHLs, predictions of new QH fluids are
made at $\,\sH=6/7,\ 10/13,\ 10/17(!),\ 13/17,\ 10/19,\ 12/19,\
14/19,\ldots\,$, and at the {\em even-denominator} fractions
$\,\sH=3/4$ and $5/8$. The CQHLs that yield even-denominator Hall
fractions have a structure that can naturally be interpreted as
describing double-layer QH systems. Furthermore, staying within these
three subclasses, we do {\em not\,} expect {\em stable(!)\,} QH
fluids to form at $\,\sH=9/11,\ 11/17,\ 14/17,\ 13/19$, and $\mini
15/19\mini$ in $\Sigma_1$, where we have omitted fractions with
$\dH\geq 21\,$ and within the ``domain of attraction'' of the most
stable Laughlin fluid at $\,\sH=1$. {\em None\mini} of these
fractions has been observed experimentally! These predictions are
rather different {}from those of the standard hierarchy
schemes~\cite{HH,JG}; for further discussions, see~\cite{FST}!
\rule[-4mm]{0mm}{5mm}

\noi
$\bullet\ $ {\em Concluding remarks.} By the bijection theorem, {\em
all\,} the statements about \Lmini\ {\em chiral\,} QH lattices in
$\,\Sigma_1^-=[1/2,1)\,$ have their precise analogues in the shifted
subwindows $\,\Sigp^-=[1/(2\mini p),1/(2\mini p-1)\mini)$, for
$p=2,3,\ldots\ $. For example, interpreting the phase transitions
observed at $\,\sH=2/3,\ 3/5$, and possibly $5/7$ in $\Sigma_1^-$ as
{\em structural phase transitions}, we predict analogous transitions
at the Hall fractions $\,\sH=2/7,\ 3/11,\ (5/17)\,$ in $\Sigma_2^-$,
and at $\,\sH=2/11,\ 3/17,\ (5/27)$ in $\Sigma_3^-$.

We remark that when acting with the shift maps $\Sm$ on the
composite, {\em mixed-chirality\,} QH lattices at $\,\sH=n/(2n-1),\
n=2,3,\ldots\,$, corresponding to the charge-conjugation picture,
we obtain {\em non-euclidean\,} QH lattices which are {\em not\,}
primitive, where primitivity has been defined at the end of
Sect.\,2\mini. Thus, the corresponding images do not figure in the
present paper which is restricted to primitive CQHLs. A discussion
of non-primitive QH lattices and the corresponding QH fluids has
been given in~\cite[Appendix E]{FST}. In particular, in~\cite{FST},
we have described composite, non-euclidean (but ``factorized'') QH
lattices corresponding to the hierarchy fluids in the windows
$\,\Sigp^-,\ p=1,2,3$.

For the complementary subwindows $\,\Sigp^+=[1/(2\mini p+1),1/(2\mini
p)\mini)$, the one-to-one correspondences between the different
sets of \Lmini\ CQHLs are implicit in the classification result of
the uniqueness theorem discussed above.

These results may suffice to convince the reader that there is a
significant {\em structural asymmetry\,} between the sets of QH
fluids with Hall conductivities in the two complementary subwindows
$\Sigp^+$ and $\Sigp^-$, for a given $\,p=1,2,\ldots\ $, while there
is a {\em structural similarity\,} between all the sets of QH fluids
with conductivities in the ``$+$''-windows $\Sigp^+$ and all the ones
with conductivities in the ``$-$''-windows $\Sigp^-$., for
different values of $p$.

We conclude by noting that, after more then ten years since its
discovery~\cite{QHE}, the fractional QH effect in the interval
$\,0<\sH\leq 1\,$ is still an interesting field of experimental and
theoretical research. In the present paper and in~\cite{FST}, we
have argued that the QH-lattice approach provides an efficient
instrument for describing universal properties of QH fluids. In our
analysis of {\em physically relevant\,} QH lattices we have assumed
chirality and \Lminiy\ as basic properties. New and refined
experimental data in the neighbourhood of the various plateau-values
discussed above would either support or question these hypotheses,
and hence could lead to further progress in the understanding of
this fascinating effect.

\vspace{10mm}
\begin{flushleft}
{\Large\bf Acknowledgements}
\end{flushleft}

\noi
We wish to thank our colleagues Yosi Avron, Rudolf Morf, Duncan
Haldane, and Paul Wiegmann for helpful discussions and
encouragement. We also thank Mansour Shayegan for sending us a copy
of~\cite{Su} prior to publication. T.K.\ acknowledges partial
support {}from the NSF (grant DMS-9305715), and U.M.S.\
acknowledges support {}from the Onderzoeksfonds K.U.~Leuven (grant
OT/92/9).



\end{document}